\documentclass[aps,prmaterials,twocolumn,amsmath,amssymb,superscriptaddress,citeautoscript,longbibliography,floatfix]{revtex4-2}

\usepackage{graphicx}% Include figure files
\usepackage{dcolumn}% Align table columns on decimal point
\usepackage{bm}
\usepackage{color}
\usepackage{mathptmx}
\usepackage{mathrsfs}
\usepackage{multirow}
\usepackage{natbib}

\usepackage{epsf}
\usepackage{epsfig}
\usepackage{epstopdf}

\begin{document}

\title{Improved magnetostructural and magnetocaloric reversibility in magnetic Ni-Mn-In shape-memory Heusler alloy by optimizing the geometric compatibility condition}

\author{P.\ Devi}\email{Parul.Devi@cpfs.mpg.de}
\affiliation{Max Planck Institute for Chemical Physics of Solids,  N\"{o}thnitzer Str.\ 40, 01187 Dresden, Germany}

\author{C.\ Salazar Mej\'{i}a}
\affiliation{Dresden High Magnetic Field Laboratory (HLD-EMFL), Helmholtz-Zentrum Dresden-Rossendorf, 01328 Dresden, Germany}

\author{M.\ Ghorbani Zavareh}
\affiliation{Max Planck Institute for Chemical Physics of Solids,  N\"{o}thnitzer Str.\ 40, 01187 Dresden, Germany}

\author{K. K. Dubey}
\affiliation{School of Materials Science and Technology, Indian Institute of Technology (BHU), Varanasi-221005, India}

\author{Pallavi Kushwaha}
\affiliation{Max Planck Institute for Chemical Physics of Solids,  N\"{o}thnitzer Str.\ 40, 01187 Dresden, Germany}

\author{Y. Skourski}
\affiliation{Dresden High Magnetic Field Laboratory (HLD-EMFL), Helmholtz-Zentrum Dresden-Rossendorf, 01328 Dresden, Germany}

\author{C.\ Felser}
\affiliation{Max Planck Institute for Chemical Physics of Solids,  N\"{o}thnitzer Str.\ 40, 01187 Dresden, Germany}

\author{M.\ Nicklas}
\affiliation{Max Planck Institute for Chemical Physics of Solids,  N\"{o}thnitzer Str.\ 40, 01187 Dresden, Germany}

\author{Sanjay Singh}\email{ssingh.mst@iitbhu.ac.in}
\affiliation{Max Planck Institute for Chemical Physics of Solids,  N\"{o}thnitzer Str.\ 40, 01187 Dresden, Germany}
\affiliation{School of Materials Science and Technology, Indian Institute of Technology (BHU), Varanasi-221005, India}

\date{\today}

\begin{abstract}
We report an improved reversibility of magnetostriction and inverse magnetocaloric effect (MCE) for the magnetic shape-memory Heusler alloy Ni$_{1.8}$Mn$_{1.8}$In$_{0.4}$. We show that the magnetostriction and MCE crucially depends on the geometrical compatibility of the austenite and martensite phases. Detailed information on the compatibility of both phases has been obtained from the transformation matrix calculated from x-ray diffraction data. The uniqueness of the lattice parameters results in an improved reversibility of the magnetostriction and the MCE. In the thermal hysteresis region of the martensitic transformation, the maximum relative length change is 0.3\% and the adiabatic temperature change $\Delta T_{ad}\approx -10$~K in pulsed magnetic fields. Our results reveal that the approach of geometric compatibility will allow {\color{black}one} to design materials with reversible magnetostriction and reversible inverse MCE at a first-order magnetostructural phase transition in shape-memory Heusler alloys.
\end{abstract}

\maketitle

%\section{Introduction}
First-order phase transitions in magnetic materials have gained strong interest, due to their potential applicability in magnetic refrigeration at room temperature \cite{Franco2012}. Magnetic refrigeration is based on the magnetocaloric effect (MCE), which is defined as heating or cooling of a magnetic material in a changing magnetic field \cite{Bruck2005,Taubel2018}. The MCE is determined quantitatively in terms of the isothermal entropy change or the adiabatic temperature change \cite{Taubel2018}. A giant MCE around room temperature was first discovered in Gd$_{5}$(Si$_{1-x}$Ge$_{x}$)$_4$ \cite{Moore2009, Pecharsky1997}. Following this discovery, few other systems, such as LaFe$_{13-x}$Si$_{x}$ \cite{Fujita2003, Lyubina2008}, MnAs$_{1-x}$Sb$_{x}$ \cite{Wada2002}, MnFeP$_{1-x}$As$_{x}$ \cite{Tegus2002}, were found to exhibit a giant MCE.

Recently, a considerable attention has been paid to the Mn rich Ni$_{2-x}$Mn$_{1+x}$\textit{Z} ($Z = {\rm Sn}$, In, Sb) based magnetic shape-memory Heusler alloys that undergo a first-order diffusion-less martensitic phase transformation from a high-temperature high-symmetry cubic austenite phase to a low-temperature low-symmetry martensitic phase, which can have tetragonal, orthorhombic, or monoclinic symmetry \cite{Sutou2004, Liu2012}. The first-order martensitic phase transition, driven by nucleation and growth of the austenite phase, contributes to several fascinating properties, such as shape-memory, magnetic-superelasticity and caloric effects \cite{Liu2012, Zavareh2015, Krenke2005, Planes2009}. The origin of these physical properties is coming from the strong interrelation between crystal structure and magnetism. Especially, the crystallographic change at the martensitic transition can generate a large MCE useable in cooling applications \cite{Devi2018, Liu2012}.

Among the Ni$_{2-x}$Mn$_{1+x}$\textit{Z} ($Z = {\rm Sn}$, In, Sb) Heusler alloys, the In-based ones are the most promising in terms of magnetic refrigeration because of the significant cooling effect with {\color{black} reversible} adiabatic temperature change, {\color{black}$|\Delta T_{ad}|$} of up to 5.4~K \cite{Zavareh2015, Gottschall2015}. However, the large $\Delta T_{ad}$  cannot be observed in successive field cycles due to the large thermal hysteresis in these materials \cite{Zavareh2015, Liu2012}. The thermal hysteresis arises because of the lattice mismatch between austenite and martensite phases \cite{Bhattacharya2003, Devi2018a}. Therefore, nowadays most of the efforts are devoted to reduce the hysteresis aiming at a reversible MCE. The thermal hysteresis can be reduced by different methods, such as chemical pressure by doping of an appropriate element, tuning the composition, annealing conditions, or physical pressure. As a result interatomic distances change which leads to a modification of magnetic interactions {\color{black}\cite{Caron2017, Singh2016, Pons2008, Barman2008}}. {\color{black}However}, the methods used for a reduction of the hysteresis affect not only the thermal hysteresis, but also the magnetic properties which include transition temperatures, sharpness of the transition and, thus, the magnetocaloric properties. That makes the implementation of this promising strategy tricky in Heusler alloys \cite{Khovaylo2010, Khovaylo2009, Gottschall2016}.

In Heusler alloys the hysteresis is correlated with the compatibility of austenite/martensite interfaces and the compatibility itself to the reversibility of MCE \cite{Bhattacharya2003, Zhang2009}. Recently, Song \textit{et al.\ }\cite{Song2013} have shown that the hysteresis can be reduced in nonmagnetic alloys by improving the compatibility condition between austenite and martensite phases. The compatibility condition depends on the crystal structure of the martensite phase \cite{Bhattacharya2003}.
It is the {\color{black}most} simple for a cubic austenitic and a tetragonal martensitic phase. In this case the compatibility condition reduces to the constrain of a volume conserving martensitic transition \cite{Bhattacharya2003}. That is for example fulfilled in the magnetic shape-memory Heusler alloy Ni$_{2.2}$Mn$_{0.8}$Ga which exhibits a conventional, reversible MCE because of the compatibility of cubic austenite to tetragonal martensite structure \cite{Devi2018a}. However, brittleness and the low {\color{black}$\Delta T_{ad}$} value, typical for a conventional MCE hinder its technological application, motivating the search for novel materials showing an inverse MCE, exhibiting a larger cooling effect as well as providing better mechanical properties \cite{Taubel2018, Liu2012, Zavareh2015, Caron2011}.

In the present Rapid Communication, we have explored the validity of the geometric compatibility condition on the reversibility of the inverse MCE in the magnetic shape-memory Heusler alloy Ni$_{1.8}$Mn$_{1.8}$In$_{0.4}$. Following the work of Khovaylo {\it et al.\ }(Ref.\ \onlinecite{Khovaylo2009} and references therein), our starting point was the magnetic shape-memory Heusler materials belonging to the Ni$_{2}$Mn$_{1+x}$In$_{1-x}$ family, which exhibit a small thermal hysteresis of 15.5 K at $x = 0.4$ and an \textit{irreversible} adiabatic temperature change of 7~K in a pulsed magnetic field of 20~T \cite{Ito2007, Zavareh2015}. The irreversible behavior of $\Delta T_{ad}$ may {\color{black}arise} due to the large deviation of the compatibility condition of 5.7\% from unity {\color{black}which we calculate by using the lattice parameters} between austenite and martensite phases \cite{Khovaylo2009, Oikawa2006}. By tuning the ratio of the valence electron per atom, we obtained the lowest hysteresis of 9.5~K in the Ni-Mn-In family in off-stoichiometric  Ni$_{1.8}$Mn$_{1.8}$In$_{0.4}$. For this compound, we find an improved compatibility condition with only 0.49\% deviation from unity, which results in a large and \textit{reversible} behavior of magnetostriction and inverse MCE in the hysteresis region under subsequent magnetic-field cycling.

A polycrystalline ingot of Ni$_{1.8}$Mn$_{1.8}$In$_{0.4}$ was prepared by arc-melting and annealed for 3 days at $900^{\circ}$C, followed by quenching in an ice-water mixture. To collect the synchrotron x-ray powder diffraction (SXRPD) data, part of the annealed ingot was grounded into powder and further annealed at $700^{\circ}$C for 10~hours to remove the stress induced during grinding \cite{Singh2015, Singh2017}. SXRPD patterns were collected by using a wavelength of {0.20712 \AA}, at P02 beamline in Petra III, Hamburg, Germany. The magnetization measurements were investigated utilizing a Magnetic Property Measurement System (Quantum Design). Isothermal magnetic measurements $M(H)$ were measured in a Physical Property Measurement System (Quantum Design) up to 14~T. Pulsed magnetic field measurements were performed at the Dresden High Magnetic Field Laboratory (HLD), Germany, using a home-built set up. The magnetostriction experiments were carried out using a resistive strain gauge glued to the sample {\color{black}and applying 100~ms magnetic pulses.} The MCE was determined by measuring the adiabatic temperature change directly by a copper-constantan thermocouple squeezed between two pieces of sample {\color{black}in applied magnetic field pulses of approx.\ 75~ms.} The target temperature was recorded by a resistive Cernox thermometer (Lake Shore Cryotronics), {\color{black}as described in Ref. [\onlinecite{Zavareh2015}].}

\begin{figure}[t!]
\includegraphics[width=0.9\linewidth]{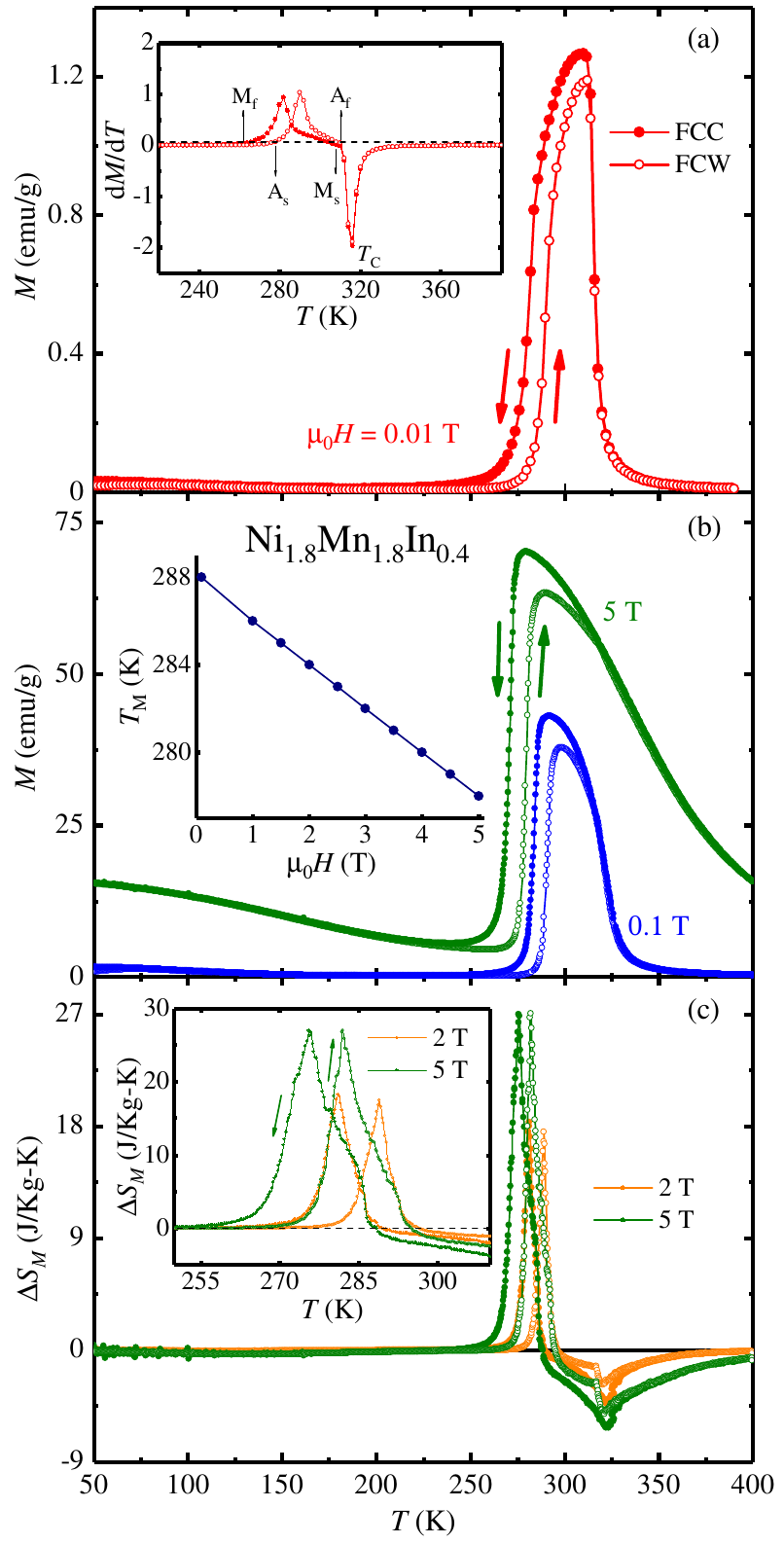}
\centering
\caption{(a) Field-cooled cooling (FCC) and field-cooled warming (FCW) magnetization $M(T)$ curves at 0.01~T for Ni$_{1.8}$Mn$_{1.8}$In$_{0.4}$. $dM(T)/dT$ is shown in the inset. (b) $M(T)$ curves at different magnetic fields of 0.1 and 5~T. The inset shows the shift of martensitic transition temperature $T_M$ {\color{black}(martensitic transition temperature upon cooling from austenitic to martensitic phase)} as function of the magnetic field. (c) Isothermal magnetic entropy change $\Delta S_{M}(T)$ calculated from the corresponding $M(T)$ curves upon cooling and heating. The inset presents $\Delta S_{M}(T)$ on an expanded scale around the martensitic transition. FCC and FCW data are represented by solid and open symbols, respectively.}
\label{Magnetization}
\end{figure}

%\sectionResult and discussion: -
Upon cooling, Ni$_{1.8}$Mn$_{1.8}$In$_{0.4}$ undergoes a direct transformation from the {\color{black}austenitic} to the {\color{black}martensitic} phase, whereas, upon heating, the reverse transformation from the martensitic to the austenitic phase takes place. Figure \ref{Magnetization}(a) displays the temperature dependence of the magnetization $M(T)$ in a magnetic field of 0.01~T following field-cooled cooling (FCC) and field-cooled warming (FCW) protocols. The inset shows the first derivative of the magnetization with respect to temperature $dM(T)/dT$, which was used to determine the characteristic temperatures: the martensitic start $M_{s} =$ 307~K and the martensitic finish temperature $M_{f} = 260$~K upon cooling and the austenitic start $A_{s} = 277$~K and the austenitic finish temperature $A_{f} = 309$~K upon warming and the Curie temperature $T_\text{C}\approx 316$~K. The width of the thermal hysteresis {\color{black} obtained from the characteristic temperatures by using the formula \Big[$\frac{({A_s{+}A_f}){-}({M_s{+}M_f})}{2}$\Big]  is 9.5~K}, which is considerably smaller than reported for the parent compound Ni$_{2}$Mn$_{1.4}$In$_{0.6}$ \cite{Khovaylo2010, Zavareh2015}.

The smaller thermal hysteresis in Ni$_{1.8}$Mn$_{1.8}$In$_{0.4}$ compared with the parent compound Ni$_{2}$Mn$_{1.4}$In$_{0.6}$ suggests a possible higher value of $\Delta S_{M}$ \cite{Khovaylo2010,Krenke2007}. To calculate the isothermal entropy change, $M(T)$ curves were measured at several magnetic fields ranging from 0.1 to 5~T following FCC and FCW protocols. Data for two representative fields are shown in Fig.\ \ref{Magnetization}(b). {\color{black}The inset of Fig.\ \ref{Magnetization} (b) shows that} increasing the applied magnetic field shifts the martensitic transition toward lower temperatures, indicating that magnetic field stabilizes the austenitic phase. $\Delta S_{M}$ was then calculated from the $M(T)$ curves by using the Maxwell relation \cite{Khovaylo2010}:
\begin{equation}
\left(\frac{\partial S}{\partial H}\right)_{T} = \left(\frac{\partial M}{\partial T}\right)_{H}
\end{equation}
The obtained isothermal-entropy change, for both heating and cooling protocols, for magnetic field changes of 2 and 5~T is shown in Fig.\ \ref{Magnetization}(c). The maximum value of $\Delta S_{M}$ calculated from $M(T)$ upon both heating and cooling is almost the same, as exemplified in the magnified view in the inset of Fig.\ \ref{Magnetization}(c). This suggests that, due to small thermal hysteresis and, consequently, a similar value of $\Delta S_{M}$ for heating and cooling protocols, Ni$_{1.8}$Mn$_{1.8}$In$_{0.4}$ possesses compatible austenite and martensite phases \cite{Devi2018a,Bhattacharya2003, Song2013}.

\begin{figure}[t!]
\includegraphics[width=0.9\linewidth]{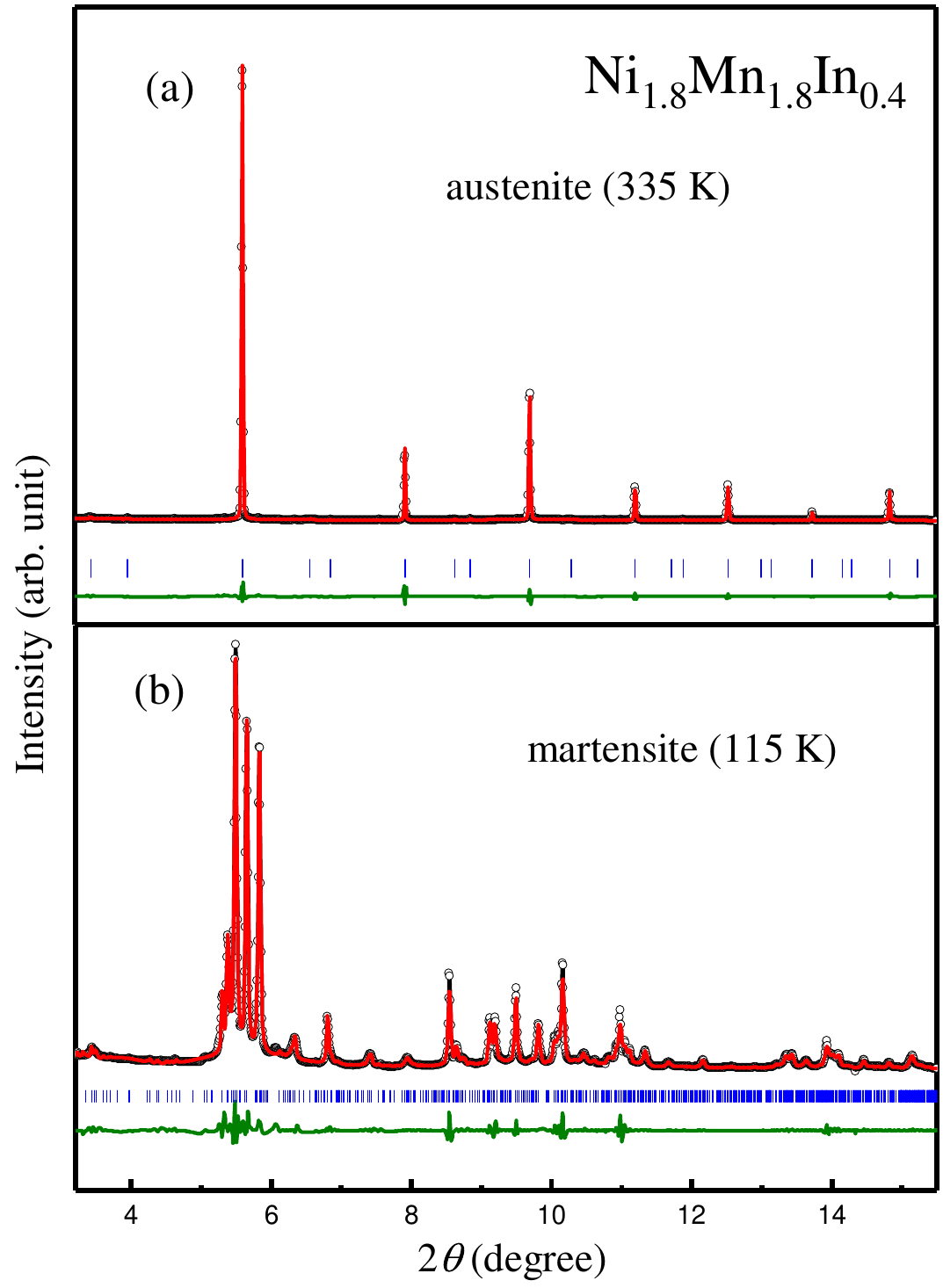}
\centering
\caption{ Synchrotron x-ray powder diffractogram of Ni$_{1.8}$Mn$_{1.8}$In$_{0.4}$ in the (a) austenite and (b) martensite phases. The experimental data, fitted curves, and residues are shown by black circles, red and green lines, respectively. The blue ticks represent the Bragg-peak positions.}
\label{XRD}
\end{figure}

Motivated by small thermal hysteresis observed in the analysis of the isothermal-entropy change, we calculated the compatibility condition for Ni$_{1.8}$Mn$_{1.8}$In$_{0.4}$ from the unit cell parameters of the austenite and martensite phases. Lebail refinements for the austenite and martensite phases are displayed in upper and lower panel of Fig.\ \ref{XRD}. The austenitic phase has a cubic structure (space group $Fm$-3$m$) with cell parameter $a_0=6.00482$~\AA\ [see Fig.\ \ref{XRD}(a)]. At 115~K, the martensitic phase has significantly more Bragg reflections [see Fig.\ \ref{XRD}(b)]. For Ni-Mn based Heusler alloys, these types of complicated diffraction patterns have been reported as modulated structures \cite{Devi2018, Singh2017}. Therefore, we further analyze the diffraction pattern taking into account both main and satellite reflections. A careful analysis of all Bragg reflections present in the martensite phase of Ni$_{1.8}$Mn$_{1.8}$In$_{0.4}$ shows that it has a 3$M$ modulated monoclinic structure (space group $P$2/$m$) with refined lattice parameters $a = 4.4359$~\AA, $b =$ 5.5684~\AA, $c = 13.0283$~\AA, and $\beta=94.0301^{\circ}$.

The compatibility condition, also known as cofactor condition, for a modulated monoclinic structure is different and more complicated in comparison with the tetragonal martensitic structure because of the existence of 12 correspondence variants of the modulated monoclinic structure whereas there are only 3 correspondence variants in the tetragonal structure \cite{Devi2018a, Bhattacharya2003}. {\color{black}However}, all of these correspondence variants have the same eigenenergy, eigenvalues, and volume change. Therefore, we consider only one of the correspondence variant for modulated monoclinic structure here. The correspondence variant {\color{black}which is also known as the transformation or lattice deformation matrix} along $\langle100\rangle$ is described as follows:
\begin{equation}
\mathbf{U}_{1}=
\begin{pmatrix}
\tau & \sigma & 0\\
\sigma & \rho & 0\\
0 & 0 & {\color{black}\chi}
\end{pmatrix}\mathrm{.}
\label{U1}
\end{equation}
Here the matrix elements are defined as:
\begin{align}
\tau &= \frac{\alpha^2+\gamma^2+2\alpha\gamma{(\sin{\color{black}\beta}-\cos{\color{black}\beta)}}}{2\sqrt{\alpha^2+\gamma^2+2\alpha\gamma\sin{\color{black}\beta}}},\\
\rho &= \frac{\alpha^2+\gamma^2+2\alpha\gamma{(\sin{\color{black}\beta}+\cos{\color{black}\beta})}}{2\sqrt{\alpha^2+\gamma^2+2\alpha\gamma\sin{\color{black}\beta}}},\\
\sigma &= \frac{\alpha^2-\gamma^2}{2\sqrt{\alpha^2+\gamma^2+2\alpha\gamma\sin{\color{black}\beta}}},{\color{black}\mathrm{~and}}\\
{\color{black}\chi} &{\color{black}= \frac{b}{a_0},}
\end{align}
{\color{black}with} $\alpha = \frac{\sqrt{2}a}{a_0}$ and $\gamma= \frac{\sqrt{2}c}{Na_{0}}$ \cite{James2000, Hane1999}.
The cubic lattice parameter is denoted as $a_{0}$ whereas, monoclinic unit cell parameters are denoted by $a, b, c$, and angle {$\beta$}. $N$ is the degree of modulation.
Thus, the transformation matrix (Eq.\ \ref{U1}) of Ni$_{1.8}$Mn$_{1.8}$In$_{0.4}$ is:
\begin{equation}
\mathbf{U}_{1}=
\begin{pmatrix}
1.0694 & 0.0109 & 0\\
0.0109 & 0.9967 & 0\\
0 & 0 & 0.9273
\end{pmatrix}
\label{U2}
\end{equation}
The determinant of the matrix results in a value of 0.9884 and its eigenvalues are 0.9273, 0.9951, and 1.0711. The middle eigenvalue is 0.9951, which is close to one ($|\lambda_{2}-1|=0.0049$) with only 0.49\% deviation from unity. Therefore, Ni$_{1.8}$Mn$_{1.8}$In$_{0.4}$ may follow the expectations for the geometric compatibility condition \cite{Bhattacharya2003, Song2013}.

\begin{figure}[t!]
\includegraphics[width=0.9\linewidth]{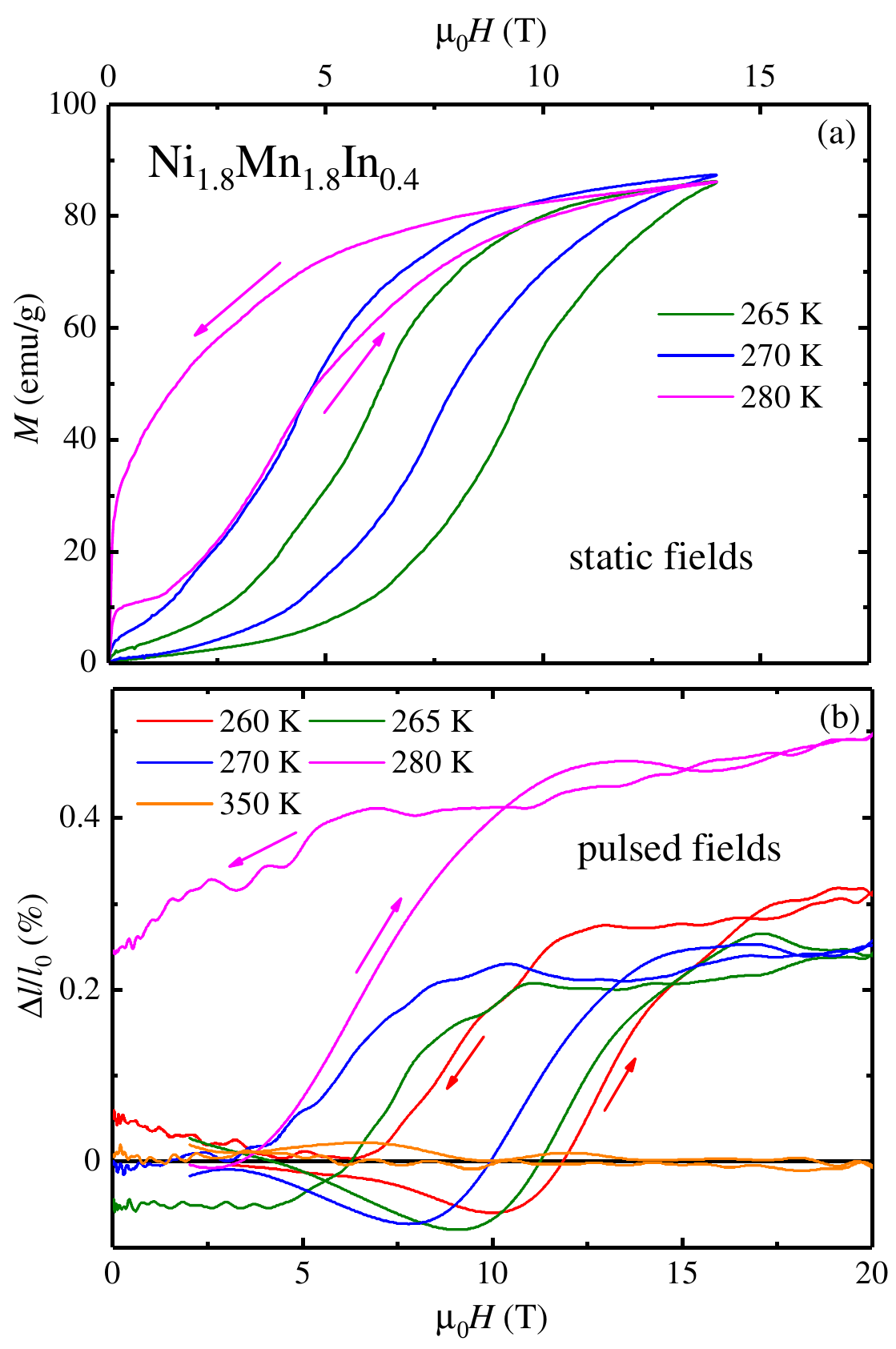}
\centering
\caption{  (a) Isothermal magnetization data $M(H)$ measured in static magnetic fields up to 14~T for temperatures below and close to $A_{s}$. (b) The relative length change $\Delta l/l_0(H)$ recorded in pulsed magnetic field experiments at different temperatures.}
\label{M(H)}
\end{figure}

To study the importance of the almost perfectly fulfilled compatibility condition on the MCE, we investigated the reversibility of field-induced magnetostriction and determined the adiabatic temperature change by direct measurements in pulsed magnetic field, accompanied by isothermal $M(H)$ recordings. Figure \ref{M(H)}(a) displays $M(H)$ isotherms at different temperatures up to 14~T. Each $M(H)$ curve was taken in the following protocol: the sample was first heated up to the fully austenitic phase and then cooled down to the fully martensitic phase followed by heating to the target temperature for the experiment $T_{i}$. By following this protocol, we assure that the sample state is not influenced by the history of measurements. As can be seen in Fig.\ \ref{M(H)}a, Ni$_{1.8}$Mn$_{1.8}$In$_{0.4}$ exhibits a field-induced reverse martensitic transition, as commonly found in Ni-Mn based shape-memory Heusler alloys \cite{Zavareh2015,Nayak2014}. For temperatures close to $A_{s}$, a magnetic field of 14~T is sufficient to induce the reverse martensitic transition.

%Magnetostriction measurement in pulsed magnetic field:
We collected magnetostriction data at different temperatures between 260 and 350~K in pulsed magnetic fields using 20~T pulses which are high enough to transform the sample to the fully austenitic phase. The results are shown in Fig.\ \ref{M(H)}(b). Each measurement was preceded by the temperature profile described above. The relative length change is determined as $\Delta l/l_0=(l-l_0)/l_0$, where $l_{0}$ is the length of the sample before each magnetic-field pulse.
At 350~K the sample is in the fully austenitic phase and no significant change in the sample length is observed when the field is applied. At temperatures below $A_{s}$, we show data for 260, 265, and 270~K, the magnetic field induces the transition from martensite to austenite. Initially, the sample compresses and then expands up to a relative length change of about 0.3\%, {\it i.e.}\ the austenitic possesses a larger volume than the martensitic phase. The effect is reversible and the size is comparable to that in other Ni-Mn-based magnetic shape-memory Heusler alloys \cite{Kainuma2006, Salazar2017, Li2010}. {\color{black}The origin of the initial compression is not fully understood, however, it has been observed in other Heusler alloys too \cite{Lazpita2016,Salazar2017}. Presumably, it is related to some re-orientation process of the martensitic variants as a self-accommodation process in preparation to the structural transition.}
At $T_{i}=280~{\rm K}>A_{s}$, part of the sample is already in the austenitic phase, as it is also visible in the $M(H)$ curve at 280~K. Due to this, the {\color{black}sample} does not exhibit an initial compression and displays a larger relative change in the sample length. On the other hand, since 280~K is within the hysteresis region {$M_{f},A_s<T_{i}<M_s,A_f$}, the transition is induced by field but, after the field is removed the sample does not transform back to a completely martensitic phase. As a consequence, a remanent expansion of about 0.2\% remains.

\begin{figure}[t!]
\includegraphics[width=0.9\linewidth]{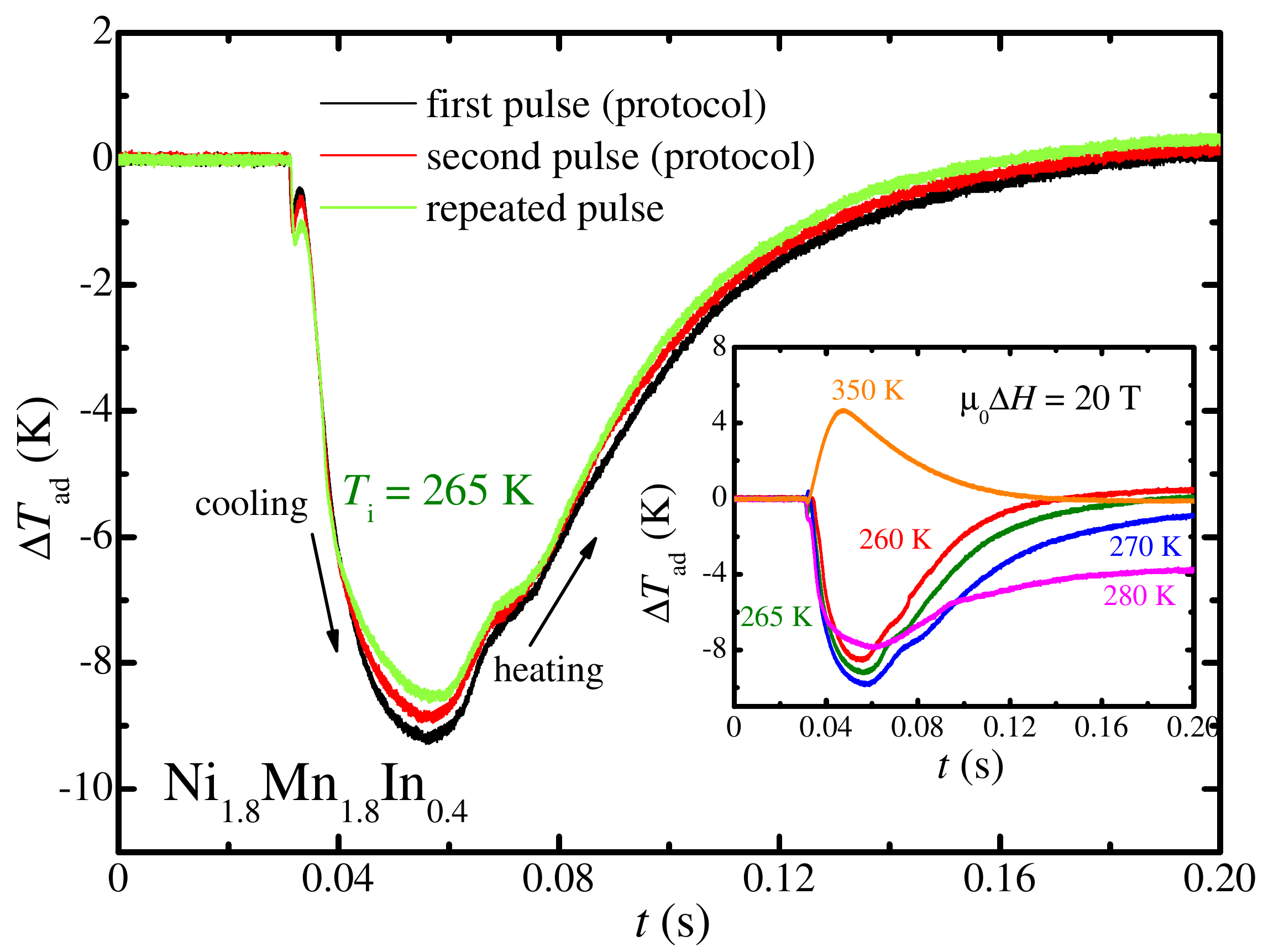}
\centering
\caption{ Time dependence of the adiabatic temperature change $\Delta T_{ad}(t)$ recorded at 265~K for a magnetic-field pulse of 20~T recorded for three different pulses. The inset shows $\Delta T_{ad}(t)$ for different sample temperatures. Each measurement (except for the repeated pulse) was preceded by heating up the sample to the fully austenitic phase followed by cooling down to the completely martensitic phase before approaching the measurement temperature. The repeated pulse was taken one hour after the second pulse.}
\label{Pulse measurement}
\end{figure}

%Direct adiabatic temperature changes in pulsed magnetic field:
In order to characterize the MCE, we detected the direct adiabatic temperature change in pulsed magnetic-fields up to 20~T (see Fig.\ \ref{Pulse measurement}). This provides us a comprehensive access to the dynamic properties of the MCE. Note, we used the same measurement protocol as before. We find indeed a reversible behavior of the inverse MCE at 260 and 265~K in the lower part of hysteresis region. Figure \ref{Pulse measurement} shows $\Delta T_{ad}(t)$ at $T_{i}=265$~K for {\color{black}three 20~T magnetic field pusles. The sample temperature before the first and second pulse was reached by overheating to the austenite phase and undercooling to the martensite phase to avoid the mixed martensitic/austenite state. The "repeated" pulse was taken one hour after the second pulse, without cooling or heating the sample in between, in order to confirm  the reversibility at 265~K. The waiting time before the pulses is determined by the cooling time of the coil used to generate the pulsed fields. 265~K is inside} the hysteresis region. However, $\Delta T_{ad}(t)$  is completely reproducible. {\color{black}As shown in Fig.\ \ref{Pulse measurement}, initially the samples cools down by around $-9$~K due to the field sweep up (time regime from 3 to 6~ms). The sample warms on the drop off side of the pulse, reaching the initial temperature again, indicating reversible behavior.} For temperatures closer to, or higher than $A_s$, the initial state cannot be reached anymore when the magnetic field is removed and the sample ends up in a mixed state which is magnetically different from the initial one. Consequently, the observed $\Delta T_{ad}$ is reduced as can be seen in the data at 270~K (see inset of Fig.\ \ref{Pulse measurement}). At this temperature, however, $\Delta T_{ad}(t)$ is almost reversible, up to 80\%. At 280~K the effect is irreversible, as also observed in the magnetostriction measurements. As expected, a conventional MCE is visible around the Curie temperature in the austenitic phase (see the data at 350~K, in the inset of Fig.\ \ref{Pulse measurement}). {\color{black}The sample warms up due to the rise of the field up to 20~T and then cools down again due to the drop to zero field. The sample reaches its the initial temperature, which evidences the reversibility of the MCE in this temperature region.}

%section Discussion and conclusion:
The improved reversible magnetostriction and inverse MCE in Ni$_{1.8}$Mn$_{1.8}$In$_{0.4}$ can be explained based solely on the change in the lattice parameters between austenite and martensite phases. For materials satisfying $\lambda_{2}$ = 1 at the martensitic transformation it is expected to have an exact interface, {\it i.e.}\ an invariant habit plane, between austenite and martensite phases. A lower interface energy then yields a smaller width of the thermal hysteresis and a higher reversibility of the MCE. In addition to that, the modulated structure is expected to play an important role for obtaining a large field-induced magnetostriction due to a lower twinning stress at the austenite/martensite interfaces \cite{Sutou2004}. A comparison with literature, in particular with the Ni$_{2}$Mn$_{1+x}Z_{1-x}$ ($Z ={\rm Sb}$, Sn) families of magnetic shape-memory Heusler alloys, evidences the connection of a large thermal hysteresis and strong deviations from the compatibility condition resulting in irreversible magnetostriction and irreversible inverse MCE \cite{Krenke2005, Sutou2004, Koyama2006, Khovaylo2010}.

To summarize, the compatibility condition is satisfied in the magnetic shape-memory Heusler alloy Ni$_{1.8}$Mn$_{1.8}$In$_{0.4}$, which exhibits only a small thermal hysteresis of 9.5~K. The compatibility condition, {\it i.e.}\ the middle eigenvalue of the transformation matrix deviates only by 0.49\% from unity, indicates a low interface energy between austenite and martensite phases. An improved value of the relative length change and inverse MCE, which shows  values up to $\Delta T_{ad}\approx -10$~K, was observed inside the thermal hysteresis region at 260 and 265~K. $\Delta T_{ad}$ remained constant within the measurement uncertainty in successive field pulses confirming the reversibility in this temperature range. Our study underlines the importance of the compatibility of austenite and martensite phases, also for modulated structures, in order to improve the reversibility of magnetostriction and inverse MCE in the region of their martensitic transformation. We conclude that by improving the compatibility condition a reversible conventional as well as an inverse MCE can be obtained.
\\

\begin{acknowledgments}
This work was financially supported by ERC Advanced Grant No.\ 291472 `Idea Heusler'.  We acknowledge the support of the HLD at HZDR, member of the European Magnetic Field Laboratory (EMFL). S.\ S.\ thanks Science and Engineering Research Board of India for financial support through Early Career Research Award and the award of Ramanujan Fellowship.
\end{acknowledgments}

\bibliography{MCE_NiMnIn_V12}

\begin{thebibliography}{41}%
\makeatletter
\providecommand \@ifxundefined [1]{%
 \@ifx{#1\undefined}
}%
\providecommand \@ifnum [1]{%
 \ifnum #1\expandafter \@firstoftwo
 \else \expandafter \@secondoftwo
 \fi
}%
\providecommand \@ifx [1]{%
 \ifx #1\expandafter \@firstoftwo
 \else \expandafter \@secondoftwo
 \fi
}%
\providecommand \natexlab [1]{#1}%
\providecommand \enquote  [1]{``#1''}%
\providecommand \bibnamefont  [1]{#1}%
\providecommand \bibfnamefont [1]{#1}%
\providecommand \citenamefont [1]{#1}%
\providecommand \href@noop [0]{\@secondoftwo}%
\providecommand \href [0]{\begingroup \@sanitize@url \@href}%
\providecommand \@href[1]{\@@startlink{#1}\@@href}%
\providecommand \@@href[1]{\endgroup#1\@@endlink}%
\providecommand \@sanitize@url [0]{\catcode `\\12\catcode `\$12\catcode
  `\&12\catcode `\#12\catcode `\^12\catcode `\_12\catcode `\%12\relax}%
\providecommand \@@startlink[1]{}%
\providecommand \@@endlink[0]{}%
\providecommand \url  [0]{\begingroup\@sanitize@url \@url }%
\providecommand \@url [1]{\endgroup\@href {#1}{\urlprefix }}%
\providecommand \urlprefix  [0]{URL }%
\providecommand \Eprint [0]{\href }%
\providecommand \doibase [0]{https://doi.org/}%
\providecommand \selectlanguage [0]{\@gobble}%
\providecommand \bibinfo  [0]{\@secondoftwo}%
\providecommand \bibfield  [0]{\@secondoftwo}%
\providecommand \translation [1]{[#1]}%
\providecommand \BibitemOpen [0]{}%
\providecommand \bibitemStop [0]{}%
\providecommand \bibitemNoStop [0]{.\EOS\space}%
\providecommand \EOS [0]{\spacefactor3000\relax}%
\providecommand \BibitemShut  [1]{\csname bibitem#1\endcsname}%
\let\auto@bib@innerbib\@empty
%</preamble>
\bibitem [{\citenamefont {Franco}\ \emph {et~al.}(2012)\citenamefont {Franco},
  \citenamefont {Blazquez}, \citenamefont {Ingale},\ and\ \citenamefont
  {Conde}}]{Franco2012}%
  \BibitemOpen
  \bibfield  {author} {\bibinfo {author} {\bibfnamefont {V.}~\bibnamefont
  {Franco}}, \bibinfo {author} {\bibfnamefont {J.~S.}\ \bibnamefont
  {Blazquez}}, \bibinfo {author} {\bibfnamefont {B.}~\bibnamefont {Ingale}},\
  and\ \bibinfo {author} {\bibfnamefont {A.}~\bibnamefont {Conde}},\ }\bibfield
   {title} {\bibinfo {title} {The magnetocaloric effect and magnetic
  refrigeration near room temperature: Materials and models},\ }\href@noop {}
  {\bibfield  {journal} {\bibinfo  {journal} {Annu. Rev. Mater. Res.}\ }\textbf
  {\bibinfo {volume} {42}},\ \bibinfo {pages} {305} (\bibinfo {year}
  {2012})}\BibitemShut {NoStop}%
\bibitem [{\citenamefont {Br\"{u}ck}(2005)}]{Bruck2005}%
  \BibitemOpen
  \bibfield  {author} {\bibinfo {author} {\bibfnamefont {E.}~\bibnamefont
  {Br\"{u}ck}},\ }\bibfield  {title} {\bibinfo {title} {Developments in
  magnetocaloric refrigeration},\ }\href@noop {} {\bibfield  {journal}
  {\bibinfo  {journal} {J. Phys. D: Appl. Phys.}\ }\textbf {\bibinfo {volume}
  {38}},\ \bibinfo {pages} {R381} (\bibinfo {year} {2005})}\BibitemShut
  {NoStop}%
\bibitem [{\citenamefont {Taubel}\ \emph {et~al.}(2018)\citenamefont {Taubel},
  \citenamefont {Gottschall}, \citenamefont {Fries}, \citenamefont {Riegg},
  \citenamefont {Soon}, \citenamefont {Skokov},\ and\ \citenamefont
  {Gutfleisch}}]{Taubel2018}%
  \BibitemOpen
  \bibfield  {author} {\bibinfo {author} {\bibfnamefont {A.}~\bibnamefont
  {Taubel}}, \bibinfo {author} {\bibfnamefont {T.}~\bibnamefont {Gottschall}},
  \bibinfo {author} {\bibfnamefont {M.}~\bibnamefont {Fries}}, \bibinfo
  {author} {\bibfnamefont {S.}~\bibnamefont {Riegg}}, \bibinfo {author}
  {\bibfnamefont {C.}~\bibnamefont {Soon}}, \bibinfo {author} {\bibfnamefont
  {K.~P.}\ \bibnamefont {Skokov}},\ and\ \bibinfo {author} {\bibfnamefont
  {O.}~\bibnamefont {Gutfleisch}},\ }\bibfield  {title} {\bibinfo {title} {A
  comparative study on the magnetocaloric properties of \text{Ni-Mn-X(-Co)}
  heusler alloys},\ }\href@noop {} {\bibfield  {journal} {\bibinfo  {journal}
  {Phys. Stat. Solidi B}\ }\textbf {\bibinfo {volume} {255}},\ \bibinfo {pages}
  {1700331} (\bibinfo {year} {2018})}\BibitemShut {NoStop}%
\bibitem [{\citenamefont {Moore}\ \emph {et~al.}(2009)\citenamefont {Moore},
  \citenamefont {Morrison}, \citenamefont {Perkins}, \citenamefont {Schlagel},
  \citenamefont {Lograsso}, \citenamefont {Gschneidner}, \citenamefont
  {Pecharsky},\ and\ \citenamefont {Cohen}}]{Moore2009}%
  \BibitemOpen
  \bibfield  {author} {\bibinfo {author} {\bibfnamefont {J.~D.}\ \bibnamefont
  {Moore}}, \bibinfo {author} {\bibfnamefont {K.}~\bibnamefont {Morrison}},
  \bibinfo {author} {\bibfnamefont {G.~K.}\ \bibnamefont {Perkins}}, \bibinfo
  {author} {\bibfnamefont {D.~L.}\ \bibnamefont {Schlagel}}, \bibinfo {author}
  {\bibfnamefont {T.~A.}\ \bibnamefont {Lograsso}}, \bibinfo {author}
  {\bibfnamefont {K.~A.}\ \bibnamefont {Gschneidner}}, \bibinfo {author}
  {\bibfnamefont {V.~K.}\ \bibnamefont {Pecharsky}},\ and\ \bibinfo {author}
  {\bibfnamefont {L.~F.}\ \bibnamefont {Cohen}},\ }\bibfield  {title} {\bibinfo
  {title} {Metamagnetism seeded by nanostructural features of
  single-crystalline \text{Gd$_{5}$(Si$_{2}$Ge$_{2}$)}},\ }\href@noop {}
  {\bibfield  {journal} {\bibinfo  {journal} {Adv. Mater.}\ }\textbf {\bibinfo
  {volume} {21}},\ \bibinfo {pages} {3780} (\bibinfo {year}
  {2009})}\BibitemShut {NoStop}%
\bibitem [{\citenamefont {Pecharsky}\ and\ \citenamefont
  {Gschneidner}(1997)}]{Pecharsky1997}%
  \BibitemOpen
  \bibfield  {author} {\bibinfo {author} {\bibfnamefont {V.~K.}\ \bibnamefont
  {Pecharsky}}\ and\ \bibinfo {author} {\bibfnamefont {K.~A.}\ \bibnamefont
  {Gschneidner}},\ }\bibfield  {title} {\bibinfo {title} {Giant magnetocaloric
  effect in \text{Gd$_{5}$(Si$_{2}$Ge$_{2}$)}},\ }\href@noop {} {\bibfield
  {journal} {\bibinfo  {journal} {Phys. Rev. Lett.}\ }\textbf {\bibinfo
  {volume} {78}},\ \bibinfo {pages} {4494} (\bibinfo {year}
  {1997})}\BibitemShut {NoStop}%
\bibitem [{\citenamefont {Fujita}\ \emph {et~al.}(2003)\citenamefont {Fujita},
  \citenamefont {Fujieda}, \citenamefont {Hasegawa},\ and\ \citenamefont
  {Fukamichi}}]{Fujita2003}%
  \BibitemOpen
  \bibfield  {author} {\bibinfo {author} {\bibfnamefont {A.}~\bibnamefont
  {Fujita}}, \bibinfo {author} {\bibfnamefont {S.}~\bibnamefont {Fujieda}},
  \bibinfo {author} {\bibfnamefont {Y.}~\bibnamefont {Hasegawa}},\ and\
  \bibinfo {author} {\bibfnamefont {K.}~\bibnamefont {Fukamichi}},\ }\bibfield
  {title} {\bibinfo {title} {Itinerant-electron metamagnetic transition and
  large magnetocaloric effects in \text{La(Fe$_{x}$Si$_{1-x}$)$_{13}$}
  compounds and their hybrides},\ }\href@noop {} {\bibfield  {journal}
  {\bibinfo  {journal} {Phys. Rev. B}\ }\textbf {\bibinfo {volume} {67}},\
  \bibinfo {pages} {104416} (\bibinfo {year} {2003})}\BibitemShut {NoStop}%
\bibitem [{\citenamefont {Lyubina}\ \emph {et~al.}(2008)\citenamefont
  {Lyubina}, \citenamefont {Nenkov}, \citenamefont {Schultz},\ and\
  \citenamefont {Gutfleisch}}]{Lyubina2008}%
  \BibitemOpen
  \bibfield  {author} {\bibinfo {author} {\bibfnamefont {J.}~\bibnamefont
  {Lyubina}}, \bibinfo {author} {\bibfnamefont {K.}~\bibnamefont {Nenkov}},
  \bibinfo {author} {\bibfnamefont {L.}~\bibnamefont {Schultz}},\ and\ \bibinfo
  {author} {\bibfnamefont {O.}~\bibnamefont {Gutfleisch}},\ }\bibfield  {title}
  {\bibinfo {title} {Multiple metamagnetic transitions in the magnetic
  refrigerant \text{La(Fe, Si)$_{13}$H$_{x}$}},\ }\href@noop {} {\bibfield
  {journal} {\bibinfo  {journal} {Phys. Rev. Lett.}\ }\textbf {\bibinfo
  {volume} {101}},\ \bibinfo {pages} {177203} (\bibinfo {year}
  {2008})}\BibitemShut {NoStop}%
\bibitem [{\citenamefont {Wada}\ \emph {et~al.}(2002)\citenamefont {Wada},
  \citenamefont {Taniguchi},\ and\ \citenamefont {Tanabe}}]{Wada2002}%
  \BibitemOpen
  \bibfield  {author} {\bibinfo {author} {\bibfnamefont {H.}~\bibnamefont
  {Wada}}, \bibinfo {author} {\bibfnamefont {K.}~\bibnamefont {Taniguchi}},\
  and\ \bibinfo {author} {\bibfnamefont {Y.}~\bibnamefont {Tanabe}},\
  }\bibfield  {title} {\bibinfo {title} {Extremely large magnetic entropy
  change of \text{MnAs$_{1-x}$Sb$_{x}$} near room temperature},\ }\href@noop {}
  {\bibfield  {journal} {\bibinfo  {journal} {Materials Transactions}\ }\textbf
  {\bibinfo {volume} {43}},\ \bibinfo {pages} {73} (\bibinfo {year}
  {2002})}\BibitemShut {NoStop}%
\bibitem [{\citenamefont {Tegus}\ \emph {et~al.}(2002)\citenamefont {Tegus},
  \citenamefont {Br\"{u}ck}, \citenamefont {Buschow},\ and\ \citenamefont
  {Boer}}]{Tegus2002}%
  \BibitemOpen
  \bibfield  {author} {\bibinfo {author} {\bibfnamefont {O.}~\bibnamefont
  {Tegus}}, \bibinfo {author} {\bibfnamefont {E.}~\bibnamefont {Br\"{u}ck}},
  \bibinfo {author} {\bibfnamefont {K.~H.~J.}\ \bibnamefont {Buschow}},\ and\
  \bibinfo {author} {\bibfnamefont {F.~R.~d.}\ \bibnamefont {Boer}},\
  }\bibfield  {title} {\bibinfo {title} {Transition-metal-based magnetic
  refrigerants for room-temperature applications},\ }\href@noop {} {\bibfield
  {journal} {\bibinfo  {journal} {Nature}\ }\textbf {\bibinfo {volume} {415}},\
  \bibinfo {pages} {150} (\bibinfo {year} {2002})}\BibitemShut {NoStop}%
\bibitem [{\citenamefont {Sutou}\ \emph {et~al.}(2004)\citenamefont {Sutou},
  \citenamefont {Imano}, \citenamefont {Koeda}, \citenamefont {Omori},
  \citenamefont {Kainuma}, \citenamefont {Ishida},\ and\ \citenamefont
  {Oikawa}}]{Sutou2004}%
  \BibitemOpen
  \bibfield  {author} {\bibinfo {author} {\bibfnamefont {Y.}~\bibnamefont
  {Sutou}}, \bibinfo {author} {\bibfnamefont {Y.}~\bibnamefont {Imano}},
  \bibinfo {author} {\bibfnamefont {N.}~\bibnamefont {Koeda}}, \bibinfo
  {author} {\bibfnamefont {T.}~\bibnamefont {Omori}}, \bibinfo {author}
  {\bibfnamefont {R.}~\bibnamefont {Kainuma}}, \bibinfo {author} {\bibfnamefont
  {K.}~\bibnamefont {Ishida}},\ and\ \bibinfo {author} {\bibfnamefont
  {K.}~\bibnamefont {Oikawa}},\ }\bibfield  {title} {\bibinfo {title} {Magnetic
  and martensitic transformations of \text{NiMnX(X=In,Sn,Sb)} ferromagnetic
  shape memory alloys},\ }\href@noop {} {\bibfield  {journal} {\bibinfo
  {journal} {Appl. Phys. Lett.}\ }\textbf {\bibinfo {volume} {85}},\ \bibinfo
  {pages} {4358} (\bibinfo {year} {2004})}\BibitemShut {NoStop}%
\bibitem [{\citenamefont {Liu}\ \emph {et~al.}(2012)\citenamefont {Liu},
  \citenamefont {Gottschal}, \citenamefont {Skokov}, \citenamefont {Moore},\
  and\ \citenamefont {Gutfleish}}]{Liu2012}%
  \BibitemOpen
  \bibfield  {author} {\bibinfo {author} {\bibfnamefont {J.}~\bibnamefont
  {Liu}}, \bibinfo {author} {\bibfnamefont {T.}~\bibnamefont {Gottschal}},
  \bibinfo {author} {\bibfnamefont {K.~P.}\ \bibnamefont {Skokov}}, \bibinfo
  {author} {\bibfnamefont {J.~D.}\ \bibnamefont {Moore}},\ and\ \bibinfo
  {author} {\bibfnamefont {O.}~\bibnamefont {Gutfleish}},\ }\bibfield  {title}
  {\bibinfo {title} {Giant magnetocaloric effect driven by structural
  transitions},\ }\href@noop {} {\bibfield  {journal} {\bibinfo  {journal}
  {Nat. Mater.}\ }\textbf {\bibinfo {volume} {11}},\ \bibinfo {pages} {620}
  (\bibinfo {year} {2012})}\BibitemShut {NoStop}%
\bibitem [{\citenamefont {Ghorbani~Zavareh}\ \emph {et~al.}(2015)\citenamefont
  {Ghorbani~Zavareh}, \citenamefont {Salazar~Mej\'{i}a}, \citenamefont {Nayak},
  \citenamefont {Skourski}, \citenamefont {Wosnitza}, \citenamefont {Felser},\
  and\ \citenamefont {Nicklas}}]{Zavareh2015}%
  \BibitemOpen
  \bibfield  {author} {\bibinfo {author} {\bibfnamefont {M.}~\bibnamefont
  {Ghorbani~Zavareh}}, \bibinfo {author} {\bibfnamefont {C.}~\bibnamefont
  {Salazar~Mej\'{i}a}}, \bibinfo {author} {\bibfnamefont {A.~K.}\ \bibnamefont
  {Nayak}}, \bibinfo {author} {\bibfnamefont {Y.}~\bibnamefont {Skourski}},
  \bibinfo {author} {\bibfnamefont {J.}~\bibnamefont {Wosnitza}}, \bibinfo
  {author} {\bibfnamefont {C.}~\bibnamefont {Felser}},\ and\ \bibinfo {author}
  {\bibfnamefont {M.}~\bibnamefont {Nicklas}},\ }\bibfield  {title} {\bibinfo
  {title} {Direct measurements of the magnetocaloric effect in pulsed magnetic
  fields: The example of the heusler alloy
  \text{Ni$_{50}$Mn$_{35}$In$_{15}$}},\ }\href@noop {} {\bibfield  {journal}
  {\bibinfo  {journal} {Appl. Phys. Lett.}\ }\textbf {\bibinfo {volume}
  {106}},\ \bibinfo {pages} {071904} (\bibinfo {year} {2015})}\BibitemShut
  {NoStop}%
\bibitem [{\citenamefont {Krenke}\ \emph {et~al.}(2005)\citenamefont {Krenke},
  \citenamefont {Acet}, \citenamefont {Wasermann}, \citenamefont {Moya},
  \citenamefont {Ma\~{n}osa},\ and\ \citenamefont {Planes}}]{Krenke2005}%
  \BibitemOpen
  \bibfield  {author} {\bibinfo {author} {\bibfnamefont {T.}~\bibnamefont
  {Krenke}}, \bibinfo {author} {\bibfnamefont {M.}~\bibnamefont {Acet}},
  \bibinfo {author} {\bibfnamefont {E.~F.}\ \bibnamefont {Wasermann}}, \bibinfo
  {author} {\bibfnamefont {X.}~\bibnamefont {Moya}}, \bibinfo {author}
  {\bibfnamefont {L.}~\bibnamefont {Ma\~{n}osa}},\ and\ \bibinfo {author}
  {\bibfnamefont {A.}~\bibnamefont {Planes}},\ }\bibfield  {title} {\bibinfo
  {title} {Inverse magnetocaloric effect in ferromagnetic \text{Ni-Mn-Sn}
  alloys},\ }\href@noop {} {\bibfield  {journal} {\bibinfo  {journal} {Nat.
  Mater.}\ }\textbf {\bibinfo {volume} {4}},\ \bibinfo {pages} {450} (\bibinfo
  {year} {2005})}\BibitemShut {NoStop}%
\bibitem [{\citenamefont {Planes}\ \emph {et~al.}(2009)\citenamefont {Planes},
  \citenamefont {Ma\~{n}osa},\ and\ \citenamefont {Acet}}]{Planes2009}%
  \BibitemOpen
  \bibfield  {author} {\bibinfo {author} {\bibfnamefont {A.}~\bibnamefont
  {Planes}}, \bibinfo {author} {\bibfnamefont {L.}~\bibnamefont {Ma\~{n}osa}},\
  and\ \bibinfo {author} {\bibfnamefont {M.}~\bibnamefont {Acet}},\ }\bibfield
  {title} {\bibinfo {title} {Martensitic transitions and the nature of
  ferromagnetism in the austenitic and martensitic states of \text {Ni-Mn-Sn}
  alloys},\ }\href@noop {} {\bibfield  {journal} {\bibinfo  {journal} {J.
  Phys.: Condens. Matter}\ }\textbf {\bibinfo {volume} {21}},\ \bibinfo {pages}
  {233201} (\bibinfo {year} {2009})}\BibitemShut {NoStop}%
\bibitem [{\citenamefont {Devi}\ \emph
  {et~al.}(2018{\natexlab{a}})\citenamefont {Devi}, \citenamefont {Singh},
  \citenamefont {Dutta}, \citenamefont {Manna}, \citenamefont {D$'$Souza},
  \citenamefont {Ikeda}, \citenamefont {Suard}, \citenamefont {Petricek},
  \citenamefont {Simon}, \citenamefont {Werner}, \citenamefont {Chadhov},
  \citenamefont {Parkin}, \citenamefont {Felser},\ and\ \citenamefont
  {Pandey}}]{Devi2018}%
  \BibitemOpen
  \bibfield  {author} {\bibinfo {author} {\bibfnamefont {P.}~\bibnamefont
  {Devi}}, \bibinfo {author} {\bibfnamefont {S.}~\bibnamefont {Singh}},
  \bibinfo {author} {\bibfnamefont {B.}~\bibnamefont {Dutta}}, \bibinfo
  {author} {\bibfnamefont {K.}~\bibnamefont {Manna}}, \bibinfo {author}
  {\bibfnamefont {S.~W.}\ \bibnamefont {D$'$Souza}}, \bibinfo {author}
  {\bibfnamefont {Y.}~\bibnamefont {Ikeda}}, \bibinfo {author} {\bibfnamefont
  {E.}~\bibnamefont {Suard}}, \bibinfo {author} {\bibfnamefont
  {V.}~\bibnamefont {Petricek}}, \bibinfo {author} {\bibfnamefont
  {P.}~\bibnamefont {Simon}}, \bibinfo {author} {\bibfnamefont
  {P.}~\bibnamefont {Werner}}, \bibinfo {author} {\bibfnamefont
  {S.}~\bibnamefont {Chadhov}}, \bibinfo {author} {\bibfnamefont {S.~S.~P.}\
  \bibnamefont {Parkin}}, \bibinfo {author} {\bibfnamefont {C.}~\bibnamefont
  {Felser}},\ and\ \bibinfo {author} {\bibfnamefont {D.}~\bibnamefont
  {Pandey}},\ }\bibfield  {title} {\bibinfo {title} {Adaptive modulation in the
  \text{Ni$_{2}$Mn$_{1.4}$In$_{0.6}$} magnetic shape-memory heusler alloy},\
  }\href@noop {} {\bibfield  {journal} {\bibinfo  {journal} {Phys. Rev. B.}\
  }\textbf {\bibinfo {volume} {97}},\ \bibinfo {pages} {224102} (\bibinfo
  {year} {2018}{\natexlab{a}})}\BibitemShut {NoStop}%
\bibitem [{\citenamefont {Gottschall}\ \emph {et~al.}(2015)\citenamefont
  {Gottschall}, \citenamefont {Skokov}, \citenamefont {Frincu},\ and\
  \citenamefont {Gutfleisch}}]{Gottschall2015}%
  \BibitemOpen
  \bibfield  {author} {\bibinfo {author} {\bibfnamefont {T.}~\bibnamefont
  {Gottschall}}, \bibinfo {author} {\bibfnamefont {K.~P.}\ \bibnamefont
  {Skokov}}, \bibinfo {author} {\bibfnamefont {B.}~\bibnamefont {Frincu}},\
  and\ \bibinfo {author} {\bibfnamefont {O.}~\bibnamefont {Gutfleisch}},\
  }\bibfield  {title} {\bibinfo {title} {Large reversible magnetocaloric effect
  in \text{Ni-Mn-In-Co}},\ }\href@noop {} {\bibfield  {journal} {\bibinfo
  {journal} {Appl. Phys. Lett.}\ }\textbf {\bibinfo {volume} {106}},\ \bibinfo
  {pages} {021901} (\bibinfo {year} {2015})}\BibitemShut {NoStop}%
\bibitem [{\citenamefont {Bhattacharya}(2003)}]{Bhattacharya2003}%
  \BibitemOpen
  \bibfield  {author} {\bibinfo {author} {\bibfnamefont {K.}~\bibnamefont
  {Bhattacharya}},\ }\href@noop {} {\emph {\bibinfo {title} {Microstructure of
  martensite: Why it forms and how it gives rise to the shape memory
  effect}}},\ edited by\ \bibinfo {editor} {\bibfnamefont {A.~P.}\ \bibnamefont
  {Sutton}}\ and\ \bibinfo {editor} {\bibfnamefont {R.~E.}\ \bibnamefont
  {Rudd}}\ (\bibinfo  {publisher} {Oxford Series on Materials Modelling},\
  \bibinfo {year} {2003})\BibitemShut {NoStop}%
\bibitem [{\citenamefont {Devi}\ \emph
  {et~al.}(2018{\natexlab{b}})\citenamefont {Devi}, \citenamefont {Zavareh},
  \citenamefont {Salazar~Mej\'{i}a}, \citenamefont {Hofmann}, \citenamefont
  {Albert}, \citenamefont {Felser}, \citenamefont {Nicklas},\ and\
  \citenamefont {Singh}}]{Devi2018a}%
  \BibitemOpen
  \bibfield  {author} {\bibinfo {author} {\bibfnamefont {P.}~\bibnamefont
  {Devi}}, \bibinfo {author} {\bibfnamefont {M.~G.}\ \bibnamefont {Zavareh}},
  \bibinfo {author} {\bibfnamefont {C.}~\bibnamefont {Salazar~Mej\'{i}a}},
  \bibinfo {author} {\bibfnamefont {K.}~\bibnamefont {Hofmann}}, \bibinfo
  {author} {\bibfnamefont {B.}~\bibnamefont {Albert}}, \bibinfo {author}
  {\bibfnamefont {C.}~\bibnamefont {Felser}}, \bibinfo {author} {\bibfnamefont
  {M.}~\bibnamefont {Nicklas}},\ and\ \bibinfo {author} {\bibfnamefont
  {S.}~\bibnamefont {Singh}},\ }\bibfield  {title} {\bibinfo {title}
  {Reversible adiabatic temperature change in the shape memory heusler alloy
  \text{Ni$_{2.2}$Mn$_{0.8}$Ga}: an effect of structural compatibility},\
  }\href@noop {} {\bibfield  {journal} {\bibinfo  {journal} {Phys. Rev.
  Mater.}\ }\textbf {\bibinfo {volume} {2}},\ \bibinfo {pages} {122401(R)}
  (\bibinfo {year} {2018}{\natexlab{b}})}\BibitemShut {NoStop}%
\bibitem [{\citenamefont {Caron}\ \emph {et~al.}(2017)\citenamefont {Caron},
  \citenamefont {Dutta}, \citenamefont {Devi}, \citenamefont {Zavareh},
  \citenamefont {Hickel}, \citenamefont {Cabassi}, \citenamefont {Bolzoni},
  \citenamefont {Fabbrici}, \citenamefont {Albertini}, \citenamefont {Felser},\
  and\ \citenamefont {Singh}}]{Caron2017}%
  \BibitemOpen
  \bibfield  {author} {\bibinfo {author} {\bibfnamefont {L.}~\bibnamefont
  {Caron}}, \bibinfo {author} {\bibfnamefont {B.}~\bibnamefont {Dutta}},
  \bibinfo {author} {\bibfnamefont {P.}~\bibnamefont {Devi}}, \bibinfo {author}
  {\bibfnamefont {M.~G.}\ \bibnamefont {Zavareh}}, \bibinfo {author}
  {\bibfnamefont {T.}~\bibnamefont {Hickel}}, \bibinfo {author} {\bibfnamefont
  {R.}~\bibnamefont {Cabassi}}, \bibinfo {author} {\bibfnamefont
  {F.}~\bibnamefont {Bolzoni}}, \bibinfo {author} {\bibfnamefont
  {S.}~\bibnamefont {Fabbrici}}, \bibinfo {author} {\bibfnamefont
  {F.}~\bibnamefont {Albertini}}, \bibinfo {author} {\bibfnamefont
  {C.}~\bibnamefont {Felser}},\ and\ \bibinfo {author} {\bibfnamefont
  {S.}~\bibnamefont {Singh}},\ }\bibfield  {title} {\bibinfo {title} {Effect of
  pt substitution on the magnetocrystalline anisotropy of \text{Ni$_{2}$MnGa}:
  A competition between chemistry and elasticity},\ }\href@noop {} {\bibfield
  {journal} {\bibinfo  {journal} {Phys. Rev. B}\ }\textbf {\bibinfo {volume}
  {96}},\ \bibinfo {pages} {054105} (\bibinfo {year} {2017})}\BibitemShut
  {NoStop}%
\bibitem [{\citenamefont {Singh}\ \emph {et~al.}(2016)\citenamefont {Singh},
  \citenamefont {D$'$Souza}, \citenamefont {Nayak}, \citenamefont {Caron},
  \citenamefont {Suard}, \citenamefont {Chadov},\ and\ \citenamefont
  {Felser}}]{Singh2016}%
  \BibitemOpen
  \bibfield  {author} {\bibinfo {author} {\bibfnamefont {S.}~\bibnamefont
  {Singh}}, \bibinfo {author} {\bibfnamefont {S.~W.}\ \bibnamefont
  {D$'$Souza}}, \bibinfo {author} {\bibfnamefont {J.}~\bibnamefont {Nayak}},
  \bibinfo {author} {\bibfnamefont {L.}~\bibnamefont {Caron}}, \bibinfo
  {author} {\bibfnamefont {E.}~\bibnamefont {Suard}}, \bibinfo {author}
  {\bibfnamefont {S.}~\bibnamefont {Chadov}},\ and\ \bibinfo {author}
  {\bibfnamefont {C.}~\bibnamefont {Felser}},\ }\bibfield  {title} {\bibinfo
  {title} {Effect of platinum substitution on the structural and magnetic
  properties of \text{Ni$_{2}$MnGa} ferromagnetic shape memory alloy},\
  }\href@noop {} {\bibfield  {journal} {\bibinfo  {journal} {Phys. Rev. B}\
  }\textbf {\bibinfo {volume} {93}},\ \bibinfo {pages} {134102} (\bibinfo
  {year} {2016})}\BibitemShut {NoStop}%
\bibitem [{\citenamefont {Pons}\ \emph {et~al.}(2008)\citenamefont {Pons},
  \citenamefont {Cesari}, \citenamefont {Segu\'{i}}, \citenamefont {Masdeu},\
  and\ \citenamefont {Santamarta}}]{Pons2008}%
  \BibitemOpen
  \bibfield  {author} {\bibinfo {author} {\bibfnamefont {J.}~\bibnamefont
  {Pons}}, \bibinfo {author} {\bibfnamefont {E.}~\bibnamefont {Cesari}},
  \bibinfo {author} {\bibfnamefont {C.}~\bibnamefont {Segu\'{i}}}, \bibinfo
  {author} {\bibfnamefont {F.}~\bibnamefont {Masdeu}},\ and\ \bibinfo {author}
  {\bibfnamefont {R.}~\bibnamefont {Santamarta}},\ }\bibfield  {title}
  {\bibinfo {title} {Ferromagnetic shape memory alloys: Alternatives to
  \text{Ni-Mn-Ga}},\ }\href@noop {} {\bibfield  {journal} {\bibinfo  {journal}
  {Mater. Sci. Eng., A}\ }\textbf {\bibinfo {volume} {481}},\ \bibinfo {pages}
  {57} (\bibinfo {year} {2008})}\BibitemShut {NoStop}%
\bibitem [{\citenamefont {Barman}\ \emph {et~al.}(2008)\citenamefont {Barman},
  \citenamefont {Chakrabarti}, \citenamefont {Singh}, \citenamefont {Banik},
  \citenamefont {Bhardwaj}, \citenamefont {Paulose}, \citenamefont {Chalke},
  \citenamefont {Panda}, \citenamefont {Mitra},\ and\ \citenamefont
  {Awasthi}}]{Barman2008}%
  \BibitemOpen
  \bibfield  {author} {\bibinfo {author} {\bibfnamefont {S.~R.}\ \bibnamefont
  {Barman}}, \bibinfo {author} {\bibfnamefont {A.}~\bibnamefont {Chakrabarti}},
  \bibinfo {author} {\bibfnamefont {S.}~\bibnamefont {Singh}}, \bibinfo
  {author} {\bibfnamefont {S.}~\bibnamefont {Banik}}, \bibinfo {author}
  {\bibfnamefont {S.}~\bibnamefont {Bhardwaj}}, \bibinfo {author}
  {\bibfnamefont {P.~L.}\ \bibnamefont {Paulose}}, \bibinfo {author}
  {\bibfnamefont {B.~A.}\ \bibnamefont {Chalke}}, \bibinfo {author}
  {\bibfnamefont {A.~K.}\ \bibnamefont {Panda}}, \bibinfo {author}
  {\bibfnamefont {A.}~\bibnamefont {Mitra}},\ and\ \bibinfo {author}
  {\bibfnamefont {A.~M.}\ \bibnamefont {Awasthi}},\ }\bibfield  {title}
  {\bibinfo {title} {Theoretical prediction and experimental study of a
  ferromagnetic shape memory alloy \text{Ga$_{2}$MnNi}},\ }\href@noop {}
  {\bibfield  {journal} {\bibinfo  {journal} {Phys. Rev. B}\ }\textbf {\bibinfo
  {volume} {78}},\ \bibinfo {pages} {134406} (\bibinfo {year}
  {2008})}\BibitemShut {NoStop}%
\bibitem [{\citenamefont {Khovaylo}\ \emph {et~al.}(2010)\citenamefont
  {Khovaylo}, \citenamefont {Skokov}, \citenamefont {Gutfleisch}, \citenamefont
  {Miki}, \citenamefont {Takagi}, \citenamefont {Kanomata}, \citenamefont
  {Koledov}, \citenamefont {Shavrov}, \citenamefont {Wang}, \citenamefont
  {Palacios}, \citenamefont {Bartolome},\ and\ \citenamefont
  {Burriel}}]{Khovaylo2010}%
  \BibitemOpen
  \bibfield  {author} {\bibinfo {author} {\bibfnamefont {V.~V.}\ \bibnamefont
  {Khovaylo}}, \bibinfo {author} {\bibfnamefont {K.~P.}\ \bibnamefont
  {Skokov}}, \bibinfo {author} {\bibfnamefont {O.}~\bibnamefont {Gutfleisch}},
  \bibinfo {author} {\bibfnamefont {H.}~\bibnamefont {Miki}}, \bibinfo {author}
  {\bibfnamefont {T.}~\bibnamefont {Takagi}}, \bibinfo {author} {\bibfnamefont
  {T.}~\bibnamefont {Kanomata}}, \bibinfo {author} {\bibfnamefont {V.~V.}\
  \bibnamefont {Koledov}}, \bibinfo {author} {\bibfnamefont {V.~G.}\
  \bibnamefont {Shavrov}}, \bibinfo {author} {\bibfnamefont {G.}~\bibnamefont
  {Wang}}, \bibinfo {author} {\bibfnamefont {E.}~\bibnamefont {Palacios}},
  \bibinfo {author} {\bibfnamefont {J.}~\bibnamefont {Bartolome}},\ and\
  \bibinfo {author} {\bibfnamefont {R.}~\bibnamefont {Burriel}},\ }\bibfield
  {title} {\bibinfo {title} {Peculiarities of the magnetocaloric properties in
  \text{Ni-Mn-Sn} ferromagnetic shape memory alloys},\ }\href@noop {}
  {\bibfield  {journal} {\bibinfo  {journal} {Phys. Rev. B}\ }\textbf {\bibinfo
  {volume} {81}},\ \bibinfo {pages} {214406} (\bibinfo {year}
  {2010})}\BibitemShut {NoStop}%
\bibitem [{\citenamefont {Khovaylo}\ \emph {et~al.}(2009)\citenamefont
  {Khovaylo}, \citenamefont {Kanomata}, \citenamefont {Tanaka}, \citenamefont
  {Nakashima}, \citenamefont {Amako}, \citenamefont {Kainuma}, \citenamefont
  {Umetsu}, \citenamefont {Morito},\ and\ \citenamefont {Miki}}]{Khovaylo2009}%
  \BibitemOpen
  \bibfield  {author} {\bibinfo {author} {\bibfnamefont {V.~V.}\ \bibnamefont
  {Khovaylo}}, \bibinfo {author} {\bibfnamefont {T.}~\bibnamefont {Kanomata}},
  \bibinfo {author} {\bibfnamefont {T.}~\bibnamefont {Tanaka}}, \bibinfo
  {author} {\bibfnamefont {M.}~\bibnamefont {Nakashima}}, \bibinfo {author}
  {\bibfnamefont {Y.}~\bibnamefont {Amako}}, \bibinfo {author} {\bibfnamefont
  {R.}~\bibnamefont {Kainuma}}, \bibinfo {author} {\bibfnamefont {R.~Y.}\
  \bibnamefont {Umetsu}}, \bibinfo {author} {\bibfnamefont {H.}~\bibnamefont
  {Morito}},\ and\ \bibinfo {author} {\bibfnamefont {H.}~\bibnamefont {Miki}},\
  }\bibfield  {title} {\bibinfo {title} {Magnetic properties of
  \text{Ni$_{50}$Mn$_{34.8}$In$_{15.2}$} probed by m\"{o}ssbauer
  spectroscopy},\ }\href@noop {} {\bibfield  {journal} {\bibinfo  {journal}
  {Phys. Rev. B}\ }\textbf {\bibinfo {volume} {80}},\ \bibinfo {pages} {144409}
  (\bibinfo {year} {2009})}\BibitemShut {NoStop}%
\bibitem [{\citenamefont {Gottschall}\ \emph {et~al.}(2016)\citenamefont
  {Gottschall}, \citenamefont {Skokov}, \citenamefont {Scheibel}, \citenamefont
  {Acet}, \citenamefont {Ghorbani~Zavareh}, \citenamefont {Skourski},
  \citenamefont {Wosnitza}, \citenamefont {Farle},\ and\ \citenamefont
  {Gutfleisch}}]{Gottschall2016}%
  \BibitemOpen
  \bibfield  {author} {\bibinfo {author} {\bibfnamefont {T.}~\bibnamefont
  {Gottschall}}, \bibinfo {author} {\bibfnamefont {K.~P.}\ \bibnamefont
  {Skokov}}, \bibinfo {author} {\bibfnamefont {F.}~\bibnamefont {Scheibel}},
  \bibinfo {author} {\bibfnamefont {M.}~\bibnamefont {Acet}}, \bibinfo {author}
  {\bibfnamefont {M.}~\bibnamefont {Ghorbani~Zavareh}}, \bibinfo {author}
  {\bibfnamefont {Y.}~\bibnamefont {Skourski}}, \bibinfo {author}
  {\bibfnamefont {J.}~\bibnamefont {Wosnitza}}, \bibinfo {author}
  {\bibfnamefont {M.}~\bibnamefont {Farle}},\ and\ \bibinfo {author}
  {\bibfnamefont {O.}~\bibnamefont {Gutfleisch}},\ }\bibfield  {title}
  {\bibinfo {title} {Dynamical effects of the martensitic transition in
  magnetocaloric heusler alloys from direct \text{$\Delta$Tad} measurements
  under different magnetic-field-sweep rates},\ }\href@noop {} {\bibfield
  {journal} {\bibinfo  {journal} {Phys. Rev. A}\ }\textbf {\bibinfo {volume}
  {5}},\ \bibinfo {pages} {024013} (\bibinfo {year} {2016})}\BibitemShut
  {NoStop}%
\bibitem [{\citenamefont {Zhang}\ \emph {et~al.}(2009)\citenamefont {Zhang},
  \citenamefont {James},\ and\ \citenamefont {M\"{u}ller}}]{Zhang2009}%
  \BibitemOpen
  \bibfield  {author} {\bibinfo {author} {\bibfnamefont {Z.}~\bibnamefont
  {Zhang}}, \bibinfo {author} {\bibfnamefont {R.~D.}\ \bibnamefont {James}},\
  and\ \bibinfo {author} {\bibfnamefont {S.}~\bibnamefont {M\"{u}ller}},\
  }\bibfield  {title} {\bibinfo {title} {Energy barriers and hysteresis in
  martensitic phase transformations},\ }\href@noop {} {\bibfield  {journal}
  {\bibinfo  {journal} {Acta Mater.}\ }\textbf {\bibinfo {volume} {57}},\
  \bibinfo {pages} {4332} (\bibinfo {year} {2009})}\BibitemShut {NoStop}%
\bibitem [{\citenamefont {Song}\ \emph {et~al.}(2013)\citenamefont {Song},
  \citenamefont {Chen}, \citenamefont {Dabade}, \citenamefont {Shield},\ and\
  \citenamefont {James}}]{Song2013}%
  \BibitemOpen
  \bibfield  {author} {\bibinfo {author} {\bibfnamefont {Y.}~\bibnamefont
  {Song}}, \bibinfo {author} {\bibfnamefont {X.}~\bibnamefont {Chen}}, \bibinfo
  {author} {\bibfnamefont {V.}~\bibnamefont {Dabade}}, \bibinfo {author}
  {\bibfnamefont {T.~W.}\ \bibnamefont {Shield}},\ and\ \bibinfo {author}
  {\bibfnamefont {R.~D.}\ \bibnamefont {James}},\ }\bibfield  {title} {\bibinfo
  {title} {Enhanced reversibility and unusual microstructure of a
  phase-transforming material},\ }\href@noop {} {\bibfield  {journal} {\bibinfo
   {journal} {Nature}\ }\textbf {\bibinfo {volume} {502}},\ \bibinfo {pages}
  {85} (\bibinfo {year} {2013})}\BibitemShut {NoStop}%
\bibitem [{\citenamefont {Caron}\ \emph {et~al.}(2011)\citenamefont {Caron},
  \citenamefont {Trung},\ and\ \citenamefont {Br\"{u}ck}}]{Caron2011}%
  \BibitemOpen
  \bibfield  {author} {\bibinfo {author} {\bibfnamefont {L.}~\bibnamefont
  {Caron}}, \bibinfo {author} {\bibfnamefont {N.~T.}\ \bibnamefont {Trung}},\
  and\ \bibinfo {author} {\bibfnamefont {E.}~\bibnamefont {Br\"{u}ck}},\
  }\bibfield  {title} {\bibinfo {title} {Pressure-tuned magnetocaloric effect
  in \text{Mn$_{0.93}$Cr$_{0.07}$CoGe}},\ }\href@noop {} {\bibfield  {journal}
  {\bibinfo  {journal} {Phys. Rev. B}\ }\textbf {\bibinfo {volume} {84}},\
  \bibinfo {pages} {020414} (\bibinfo {year} {2011})}\BibitemShut {NoStop}%
\bibitem [{\citenamefont {Ito}\ \emph {et~al.}(2007)\citenamefont {Ito},
  \citenamefont {Imano}, \citenamefont {Kainuma}, \citenamefont {Oikawa},\ and\
  \citenamefont {Ishida}}]{Ito2007}%
  \BibitemOpen
  \bibfield  {author} {\bibinfo {author} {\bibfnamefont {W.}~\bibnamefont
  {Ito}}, \bibinfo {author} {\bibfnamefont {Y.}~\bibnamefont {Imano}}, \bibinfo
  {author} {\bibfnamefont {R.}~\bibnamefont {Kainuma}}, \bibinfo {author}
  {\bibfnamefont {K.}~\bibnamefont {Oikawa}},\ and\ \bibinfo {author}
  {\bibfnamefont {K.}~\bibnamefont {Ishida}},\ }\bibfield  {title} {\bibinfo
  {title} {Martensitic and magnetic transformation behaviors in heusler-type
  \text{NiMnIn} and \text{NiCoMnIn} metamagnetic shape memory alloys},\
  }\href@noop {} {\bibfield  {journal} {\bibinfo  {journal} {Metall. and Mater.
  Trans. A}\ }\textbf {\bibinfo {volume} {38A}},\ \bibinfo {pages} {759}
  (\bibinfo {year} {2007})}\BibitemShut {NoStop}%
\bibitem [{\citenamefont {Oikawa}\ \emph {et~al.}(2006)\citenamefont {Oikawa},
  \citenamefont {Ito}, \citenamefont {Imano}, \citenamefont {Kainuma},
  \citenamefont {Ishida}, \citenamefont {Okamoto}, \citenamefont {Kitakami},\
  and\ \citenamefont {Kanomata}}]{Oikawa2006}%
  \BibitemOpen
  \bibfield  {author} {\bibinfo {author} {\bibfnamefont {K.}~\bibnamefont
  {Oikawa}}, \bibinfo {author} {\bibfnamefont {W.}~\bibnamefont {Ito}},
  \bibinfo {author} {\bibfnamefont {Y.}~\bibnamefont {Imano}}, \bibinfo
  {author} {\bibfnamefont {R.}~\bibnamefont {Kainuma}}, \bibinfo {author}
  {\bibfnamefont {K.}~\bibnamefont {Ishida}}, \bibinfo {author} {\bibfnamefont
  {S.}~\bibnamefont {Okamoto}}, \bibinfo {author} {\bibfnamefont
  {O.}~\bibnamefont {Kitakami}},\ and\ \bibinfo {author} {\bibfnamefont
  {T.}~\bibnamefont {Kanomata}},\ }\bibfield  {title} {\bibinfo {title} {Effect
  of magnetic field on martensitic transition of
  \text{Ni$_{46}$Mn$_{41}$In$_{13}$} heusler alloy},\ }\href@noop {} {\bibfield
   {journal} {\bibinfo  {journal} {Appl. Phys. Lett.}\ }\textbf {\bibinfo
  {volume} {88}},\ \bibinfo {pages} {122507} (\bibinfo {year}
  {2006})}\BibitemShut {NoStop}%
\bibitem [{\citenamefont {Singh}\ \emph {et~al.}(2015)\citenamefont {Singh},
  \citenamefont {Kushwaha}, \citenamefont {Scheibel}, \citenamefont {Liermann},
  \citenamefont {Barman}, \citenamefont {Acet}, \citenamefont {Felser},\ and\
  \citenamefont {Pandey}}]{Singh2015}%
  \BibitemOpen
  \bibfield  {author} {\bibinfo {author} {\bibfnamefont {S.}~\bibnamefont
  {Singh}}, \bibinfo {author} {\bibfnamefont {P.}~\bibnamefont {Kushwaha}},
  \bibinfo {author} {\bibfnamefont {F.}~\bibnamefont {Scheibel}}, \bibinfo
  {author} {\bibfnamefont {H.~P.}\ \bibnamefont {Liermann}}, \bibinfo {author}
  {\bibfnamefont {S.~R.}\ \bibnamefont {Barman}}, \bibinfo {author}
  {\bibfnamefont {M.}~\bibnamefont {Acet}}, \bibinfo {author} {\bibfnamefont
  {C.}~\bibnamefont {Felser}},\ and\ \bibinfo {author} {\bibfnamefont
  {D.}~\bibnamefont {Pandey}},\ }\bibfield  {title} {\bibinfo {title} {Residual
  stress induced stabilization of martensite phase and its effect on the
  magetostructural transition in \text{Mn}-rich \text{Ni-Mn-In/Ga} magnetic
  shape memory alloys},\ }\href@noop {} {\bibfield  {journal} {\bibinfo
  {journal} {Phys. Rev. B}\ }\textbf {\bibinfo {volume} {92}},\ \bibinfo
  {pages} {02015} (\bibinfo {year} {2015})}\BibitemShut {NoStop}%
\bibitem [{\citenamefont {Singh}\ \emph {et~al.}(2017)\citenamefont {Singh},
  \citenamefont {Dutta}, \citenamefont {D$'$Souza}, \citenamefont {Zavareh},
  \citenamefont {Devi}, \citenamefont {Gibbs}, \citenamefont {Hickel},
  \citenamefont {Chadov}, \citenamefont {Felser},\ and\ \citenamefont
  {Pandey}}]{Singh2017}%
  \BibitemOpen
  \bibfield  {author} {\bibinfo {author} {\bibfnamefont {S.}~\bibnamefont
  {Singh}}, \bibinfo {author} {\bibfnamefont {B.}~\bibnamefont {Dutta}},
  \bibinfo {author} {\bibfnamefont {S.~W.}\ \bibnamefont {D$'$Souza}}, \bibinfo
  {author} {\bibfnamefont {M.~G.}\ \bibnamefont {Zavareh}}, \bibinfo {author}
  {\bibfnamefont {P.}~\bibnamefont {Devi}}, \bibinfo {author} {\bibfnamefont
  {A.~S.}\ \bibnamefont {Gibbs}}, \bibinfo {author} {\bibfnamefont
  {T.}~\bibnamefont {Hickel}}, \bibinfo {author} {\bibfnamefont
  {S.}~\bibnamefont {Chadov}}, \bibinfo {author} {\bibfnamefont
  {C.}~\bibnamefont {Felser}},\ and\ \bibinfo {author} {\bibfnamefont
  {D.}~\bibnamefont {Pandey}},\ }\bibfield  {title} {\bibinfo {title} {Robust
  bain distortion in the premartensite phase of a platinum-substituted
  \text{Ni$_{2}$MnGa} magnetic shape memory alloy},\ }\href@noop {} {\bibfield
  {journal} {\bibinfo  {journal} {Nat. commun.}\ }\textbf {\bibinfo {volume}
  {8}},\ \bibinfo {pages} {1006} (\bibinfo {year} {2017})}\BibitemShut
  {NoStop}%
\bibitem [{\citenamefont {Krenke}\ \emph {et~al.}(2007)\citenamefont {Krenke},
  \citenamefont {Duman}, \citenamefont {Acet}, \citenamefont {Wassermann},
  \citenamefont {Moya}, \citenamefont {Ma\~{n}osa}, \citenamefont {Planes},
  \citenamefont {Suard},\ and\ \citenamefont {Ouladdiaf}}]{Krenke2007}%
  \BibitemOpen
  \bibfield  {author} {\bibinfo {author} {\bibfnamefont {T.}~\bibnamefont
  {Krenke}}, \bibinfo {author} {\bibfnamefont {E.}~\bibnamefont {Duman}},
  \bibinfo {author} {\bibfnamefont {M.}~\bibnamefont {Acet}}, \bibinfo {author}
  {\bibfnamefont {E.~F.}\ \bibnamefont {Wassermann}}, \bibinfo {author}
  {\bibfnamefont {X.}~\bibnamefont {Moya}}, \bibinfo {author} {\bibfnamefont
  {L.}~\bibnamefont {Ma\~{n}osa}}, \bibinfo {author} {\bibfnamefont
  {A.}~\bibnamefont {Planes}}, \bibinfo {author} {\bibfnamefont
  {E.}~\bibnamefont {Suard}},\ and\ \bibinfo {author} {\bibfnamefont
  {B.}~\bibnamefont {Ouladdiaf}},\ }\bibfield  {title} {\bibinfo {title}
  {Magnetic superelasticity and inverse magnetocaloric effect in \text{
  Ni-Mn-In}},\ }\href@noop {} {\bibfield  {journal} {\bibinfo  {journal} {Phys.
  Rev. B}\ }\textbf {\bibinfo {volume} {75}},\ \bibinfo {pages} {104414}
  (\bibinfo {year} {2007})}\BibitemShut {NoStop}%
\bibitem [{\citenamefont {James}\ and\ \citenamefont {Hane}(2000)}]{James2000}%
  \BibitemOpen
  \bibfield  {author} {\bibinfo {author} {\bibfnamefont {R.~D.}\ \bibnamefont
  {James}}\ and\ \bibinfo {author} {\bibfnamefont {K.~F.}\ \bibnamefont
  {Hane}},\ }\bibfield  {title} {\bibinfo {title} {Martensitic transformations
  and shape memory materials},\ }\href@noop {} {\bibfield  {journal} {\bibinfo
  {journal} {Acta Mater.}\ }\textbf {\bibinfo {volume} {48}},\ \bibinfo {pages}
  {197} (\bibinfo {year} {2000})}\BibitemShut {NoStop}%
\bibitem [{\citenamefont {Hane}\ and\ \citenamefont {James}(1999)}]{Hane1999}%
  \BibitemOpen
  \bibfield  {author} {\bibinfo {author} {\bibfnamefont {K.~F.}\ \bibnamefont
  {Hane}}\ and\ \bibinfo {author} {\bibfnamefont {R.~D.}\ \bibnamefont
  {James}},\ }\bibfield  {title} {\bibinfo {title} {Microstructure in the cubic
  to monoclinic transition in titanium-nickel shape memory alloys},\
  }\href@noop {} {\bibfield  {journal} {\bibinfo  {journal} {Acta Mater.}\
  }\textbf {\bibinfo {volume} {47}},\ \bibinfo {pages} {2603} (\bibinfo {year}
  {1999})}\BibitemShut {NoStop}%
\bibitem [{\citenamefont {Nayak}\ \emph {et~al.}(2014)\citenamefont {Nayak},
  \citenamefont {Salazar~Mej\'{i}a}, \citenamefont {D$^{\prime}$Souza},
  \citenamefont {Chadov}, \citenamefont {Skourski}, \citenamefont {Felser},\
  and\ \citenamefont {Nicklas}}]{Nayak2014}%
  \BibitemOpen
  \bibfield  {author} {\bibinfo {author} {\bibfnamefont {A.~K.}\ \bibnamefont
  {Nayak}}, \bibinfo {author} {\bibfnamefont {C.}~\bibnamefont
  {Salazar~Mej\'{i}a}}, \bibinfo {author} {\bibfnamefont {S.~W.}\ \bibnamefont
  {D$^{\prime}$Souza}}, \bibinfo {author} {\bibfnamefont {S.}~\bibnamefont
  {Chadov}}, \bibinfo {author} {\bibfnamefont {Y.}~\bibnamefont {Skourski}},
  \bibinfo {author} {\bibfnamefont {C.}~\bibnamefont {Felser}},\ and\ \bibinfo
  {author} {\bibfnamefont {M.}~\bibnamefont {Nicklas}},\ }\bibfield  {title}
  {\bibinfo {title} {Large field-induced irreversibility in {N}i-{M}n based
  {H}eusler shape-memoryalloys: A pulsed magnetic field study},\ }\href@noop {}
  {\bibfield  {journal} {\bibinfo  {journal} {Phys. Rev. B}\ }\textbf {\bibinfo
  {volume} {90}},\ \bibinfo {pages} {220408} (\bibinfo {year}
  {2014})}\BibitemShut {NoStop}%
\bibitem [{\citenamefont {Kainuma}\ \emph {et~al.}(2006)\citenamefont
  {Kainuma}, \citenamefont {Imano}, \citenamefont {Ito}, \citenamefont
  {Morito}, \citenamefont {Sutou}, \citenamefont {Oikawa}, \citenamefont
  {Fujita}, \citenamefont {Ishida}, \citenamefont {Okamoto},\ and\
  \citenamefont {Kitakami}}]{Kainuma2006}%
  \BibitemOpen
  \bibfield  {author} {\bibinfo {author} {\bibfnamefont {R.}~\bibnamefont
  {Kainuma}}, \bibinfo {author} {\bibfnamefont {Y.}~\bibnamefont {Imano}},
  \bibinfo {author} {\bibfnamefont {W.}~\bibnamefont {Ito}}, \bibinfo {author}
  {\bibfnamefont {H.}~\bibnamefont {Morito}}, \bibinfo {author} {\bibfnamefont
  {Y.}~\bibnamefont {Sutou}}, \bibinfo {author} {\bibfnamefont
  {K.}~\bibnamefont {Oikawa}}, \bibinfo {author} {\bibfnamefont
  {A.}~\bibnamefont {Fujita}}, \bibinfo {author} {\bibfnamefont
  {K.}~\bibnamefont {Ishida}}, \bibinfo {author} {\bibfnamefont
  {S.}~\bibnamefont {Okamoto}},\ and\ \bibinfo {author} {\bibfnamefont
  {K.}~\bibnamefont {Kitakami}},\ }\bibfield  {title} {\bibinfo {title}
  {Metamagnetic shape memory effect in a heusler-type
  \text{Ni$_{43}$Co$_{7}$Mn$_{39}$Sn$_{11}$} polycrystalline alloy},\
  }\href@noop {} {\bibfield  {journal} {\bibinfo  {journal} {Appl. Phys.
  Lett.}\ }\textbf {\bibinfo {volume} {88}},\ \bibinfo {pages} {192513}
  (\bibinfo {year} {2006})}\BibitemShut {NoStop}%
\bibitem [{\citenamefont {Salazar~Mej\'{i}a}\ \emph {et~al.}(2017)\citenamefont
  {Salazar~Mej\'{i}a}, \citenamefont {K\"{u}chler}, \citenamefont {Nayak},
  \citenamefont {Skourski}, \citenamefont {Wosnitza}, \citenamefont {Felser},\
  and\ \citenamefont {Nicklas}}]{Salazar2017}%
  \BibitemOpen
  \bibfield  {author} {\bibinfo {author} {\bibfnamefont {C.}~\bibnamefont
  {Salazar~Mej\'{i}a}}, \bibinfo {author} {\bibfnamefont {R.}~\bibnamefont
  {K\"{u}chler}}, \bibinfo {author} {\bibfnamefont {A.~K.}\ \bibnamefont
  {Nayak}}, \bibinfo {author} {\bibfnamefont {Y.}~\bibnamefont {Skourski}},
  \bibinfo {author} {\bibfnamefont {J.}~\bibnamefont {Wosnitza}}, \bibinfo
  {author} {\bibfnamefont {C.}~\bibnamefont {Felser}},\ and\ \bibinfo {author}
  {\bibfnamefont {M.}~\bibnamefont {Nicklas}},\ }\bibfield  {title} {\bibinfo
  {title} {Uniaxial-stress tuned large magnetic-shape-memory effect in
  \text{Ni-Co-Mn-Sb} heusler alloys},\ }\href@noop {} {\bibfield  {journal}
  {\bibinfo  {journal} {Appl. Phys. Lett.}\ }\textbf {\bibinfo {volume}
  {110}},\ \bibinfo {pages} {071901} (\bibinfo {year} {2017})}\BibitemShut
  {NoStop}%
\bibitem [{\citenamefont {Li}\ \emph {et~al.}(2010)\citenamefont {Li},
  \citenamefont {Jing}, \citenamefont {Zhang}, \citenamefont {Yu},
  \citenamefont {Chen}, \citenamefont {Kang}, \citenamefont {Cao},\ and\
  \citenamefont {Zhang}}]{Li2010}%
  \BibitemOpen
  \bibfield  {author} {\bibinfo {author} {\bibfnamefont {Z.}~\bibnamefont
  {Li}}, \bibinfo {author} {\bibfnamefont {C.}~\bibnamefont {Jing}}, \bibinfo
  {author} {\bibfnamefont {H.~L.}\ \bibnamefont {Zhang}}, \bibinfo {author}
  {\bibfnamefont {D.~H.}\ \bibnamefont {Yu}}, \bibinfo {author} {\bibfnamefont
  {L.}~\bibnamefont {Chen}}, \bibinfo {author} {\bibfnamefont {B.~J.}\
  \bibnamefont {Kang}}, \bibinfo {author} {\bibfnamefont {S.~X.}\ \bibnamefont
  {Cao}},\ and\ \bibinfo {author} {\bibfnamefont {J.~C.}\ \bibnamefont
  {Zhang}},\ }\bibfield  {title} {\bibinfo {title} {A large and reproducible
  metamagnetic shape memory effect in polycrystalline
  \text{Ni$_{45}$Co$_{5}$Mn$_{37}$In$_{13}$} heusler alloy},\ }\href@noop {}
  {\bibfield  {journal} {\bibinfo  {journal} {J. Appl. Phys.}\ }\textbf
  {\bibinfo {volume} {108}},\ \bibinfo {pages} {113908} (\bibinfo {year}
  {2010})}\BibitemShut {NoStop}%
\bibitem [{\citenamefont {Lazpita}\ \emph {et~al.}(2016)\citenamefont
  {Lazpita}, \citenamefont {Sasmaz}, \citenamefont {Cesari}, \citenamefont
  {Barandiaran}, \citenamefont {Gutierrez},\ and\ \citenamefont
  {Chernenko}}]{Lazpita2016}%
  \BibitemOpen
  \bibfield  {author} {\bibinfo {author} {\bibfnamefont {P.}~\bibnamefont
  {Lazpita}}, \bibinfo {author} {\bibfnamefont {M.}~\bibnamefont {Sasmaz}},
  \bibinfo {author} {\bibfnamefont {E.}~\bibnamefont {Cesari}}, \bibinfo
  {author} {\bibfnamefont {J.~M.}\ \bibnamefont {Barandiaran}}, \bibinfo
  {author} {\bibfnamefont {J.}~\bibnamefont {Gutierrez}},\ and\ \bibinfo
  {author} {\bibfnamefont {V.~A.}\ \bibnamefont {Chernenko}},\ }\bibfield
  {title} {\bibinfo {title} {Martensitic transformation and magnetic field
  induced effects in \text{Ni$_{42}$Co$_8$Mn$_{39}$Sn$_{11}$} metamagnetic
  shape memory alloy},\ }\href@noop {} {\bibfield  {journal} {\bibinfo
  {journal} {Acta Mater.}\ }\textbf {\bibinfo {volume} {109}},\ \bibinfo
  {pages} {170e176} (\bibinfo {year} {2016})}\BibitemShut {NoStop}%
\bibitem [{\citenamefont {Koyama}\ \emph {et~al.}(2006)\citenamefont {Koyama},
  \citenamefont {Okada}, \citenamefont {Watanabe}, \citenamefont {Kanomata},
  \citenamefont {Kainuma}, \citenamefont {Ito}, \citenamefont {Oikawa},\ and\
  \citenamefont {Ishida}}]{Koyama2006}%
  \BibitemOpen
  \bibfield  {author} {\bibinfo {author} {\bibfnamefont {K.}~\bibnamefont
  {Koyama}}, \bibinfo {author} {\bibfnamefont {H.}~\bibnamefont {Okada}},
  \bibinfo {author} {\bibfnamefont {K.}~\bibnamefont {Watanabe}}, \bibinfo
  {author} {\bibfnamefont {T.}~\bibnamefont {Kanomata}}, \bibinfo {author}
  {\bibfnamefont {R.}~\bibnamefont {Kainuma}}, \bibinfo {author} {\bibfnamefont
  {W.}~\bibnamefont {Ito}}, \bibinfo {author} {\bibfnamefont {K.}~\bibnamefont
  {Oikawa}},\ and\ \bibinfo {author} {\bibfnamefont {K.}~\bibnamefont
  {Ishida}},\ }\bibfield  {title} {\bibinfo {title} {Observation of large
  magnetoresistance of magnetic {Heusler} alloy
  \text{Ni$_{50}$Mn$_{36}$In$_{14}$} in high magnetic fields},\ }\href@noop {}
  {\bibfield  {journal} {\bibinfo  {journal} {Appl. Phys. Lett.}\ }\textbf
  {\bibinfo {volume} {89}},\ \bibinfo {pages} {182510} (\bibinfo {year}
  {2006})}\BibitemShut {NoStop}%
\end{thebibliography}%

\end{document}